\documentclass[
  journal=pasa,
  manuscript=research-paper, 
  year=2024,
  volume=XX,
]{cup-journal}

\usepackage{microtype,siunitx,booktabs,amssymb,amsmath}
\title{Spectral-line performance of low-frequency radio telescope arrays: SKA-Low stations}

\author{Lister Staveley-Smith}
\affiliation{International Centre for Radio Astronomy Research (ICRAR), University of Western Australia, 35 Stirling Highway, Crawley, WA 6009, Australia}
\email[Lister Staveley-Smith]{Lister.Staveley-Smith@uwa.edu.au}

\received {dd Mmm YYYY}
\revised  {dd Mmm YYYY}
\accepted {dd Mmm YYYY}
\published{XX Month 202X}


\begin{document}
\setlength\emergencystretch\columnwidth

\begin{abstract}
The effects of diffraction, reflection and mutual coupling on the spectral smoothness of radio telescopes becomes increasingly important at low frequencies, where the observing wavelength may be significant compared with the antenna or array dimensions. These effects are important for both traditional parabolic antennas, which are prone to the `standing wave' phenomenon caused by interference between direct and scattered wavefronts, and aperture arrays, such as the SKA-Low, MWA, HERA, and LOFAR which have more complicated scattering geometries and added dependence on pointing direction (scan angle). Electromagnetic modelling of these effects is computationally intensive and often only possible at coarse frequency resolution. Therefore, using the example of SKA-Low station configurations, we investigate the feasibility of parameterising scattering matrices, and separating antenna and array contributions to telescope chromaticity. 
This allows deeper insights into the effect on spectral smoothness and frequency-dependent beam patterns of differing antenna configurations. Even for the complicated SKA-Low element design, band-limited delay-space techniques appear to produce similar results to brute-force electromagnetic models, and allow for faster computation of station beam hypercubes (position, frequency and polarisation-dependent point spread functions) at arbitrary spectral resolution. As such techniques could facilitate improvements in the design of low-frequency spectral-line surveys, we conduct a simulated Cosmic Dawn experiment using different observing techniques and station configurations.
\end{abstract}

\section{Introduction}
\label{sec:intro}

The last two decades has seen a remarkable re-growth in low-frequency radio astronomy with the advent of new telescopes such as the Low-Frequency Array \citep[LOFAR;][]{2013A&A...556A...2V}, the Long Wavelength Array \citep[LWA;][]{2009IEEEP..97.1421E}, the Murchison Widefield Array \citep[MWA;][]{2013PASA...30....7T}, the Canadian Hydrogen Intensity Mapping Experiment \citep[CHIME; ][]{2022ApJS..261...29C}, the Hydrogen Epoch of Reionization Array \citep[HERA;][]{2017PASP..129d5001D}, Tianlai \citep{2021MNRAS.506.3455W} and the Square Kilometre Array \citep[SKA;][]{2009IEEEP..97.1482D}. This growth has principally been driven by the desire to explore the distant Universe using the highly-redshifted 21-cm line of neutral hydrogen (HI) at the epoch of early star and galaxy formation \citep{2015aska.confE...1K}, and as a tracer of cosmology and large scale structure \citep{2015ApJ...803...21B} at later epochs. This renewed interest has also given rise to exciting new low-frequency studies of Fast Radio Bursts \citep{2019Natur.566..230C}, pulsars \citep{2017ApJ...851...20M}, interplanetary scintillation \citep{2018MNRAS.474.4937C}, transient sources \citep{2022Natur.601..526H}, compact extragalactic radio sources \citep{2022MNRAS.512.5358R}, the Sun \citep{2023A&A...670A.169B}, and nearby stars \citep{2021NatAs...5.1233C}.

However, the primary aim of studying the redshifted 21-cm line at redshifts $z\gg1$ has proven unexpectedly difficult, despite these telescopes having been constructed with the requisite theoretical sensitivity. The reasons for this are variously attributed to radio-frequency interference (RFI), wide fields-of-view, spectral variance in beam parameters, strong Galactic background radiation, ionospheric Faraday rotation, and calibration difficulties. Taken in isolation, solutions to each of the above appear to exist, but the combination has so far proved intractable. Radical designs for new low-frequency telescopes (e.g. lunar orbit or lunar far-side) are being considered to eliminate at least some of the variables.  

Underlying most of the difficulties is that low-frequency antennas are typically only a small number of wavelengths across (e.g.\ 6--40 in the case of an SKA-Low station), implying that electromagnetic self-interaction and diffraction effects will be important. Compounding this is that full electromagnetic modelling of complex antenna structures is computationally difficult. For example, modelling a single SKA-Low station at a single frequency requires in excess of $10^6$ degrees of freedom \citep{2022JATIS...8a1017B}, even with substantial approximations regarding ground plane extent and amplifier impedance matching. 
However, substantial compute efficiency can be achieved by separate consideration of the strong near-field interactions within compact sub-arrays and the lesser interaction between distant sub-arrays \citep{2018ITAP...66.1805B, 2022RAA....22f5020S}. 
The approach taken in this paper is similar, in that the geometric properties of the antenna array are used define the delay, and hence the multi-frequency phase properties of the array \citep{1970ITAP...18..741W}, with the more slowly changing amplitude properties calculated from a fit or interpolation of the scattering matrix and the embedded element pattern (EEP). 

Such an approach is justified in the case of 21-cm cosmology, where the structures of interest lie at frequency scales $\Delta \nu < 20$ MHz, which arise from delays $\tau > 50$ ns. The corresponding spatial scale (15 m) is above the near-field limit of a few wavelengths where strong mutual coupling will occur. This approach considerably eases computational requirements, and therefore allows for rapid predictions of the full all-sky spectral response at almost arbitrary frequency resolution, including computation of station beams at all frequencies and positions for arbitrary array configurations. Moreover, updated models can subsequently be computed according to changes in antenna performance, antenna element layout, or masking of antennas from the array.

Characterisation of low-frequency arrays in delay space is relatively common, mainly for its usefulness in mitigating instrumental effects such as cable reflections, which result in sinusoidal spectral ripple \citep{2016ApJ...825....9T,2024arXiv240204008O}. Primary wavefront delays due to baseline length and orientation are also commonly used to separate out foregrounds and RFI. Together, such delay-delay plots are used to demarcate the so-called Epoch of Reionization (EOR) `foreground wedge' \citep{2010ApJ...724..526D,2014PhRvD..90b3018L}. 

However, full electromagnetic characterisation of the fitness-for-purpose of low-frequency arrays for challenging cosmological spectral-line experiments (e.g.\ Epoch of Reionisation, Intensity mapping, Baryonic Acoustic oscillations) is difficult. For HERA, \citet{2024arXiv240608549R} report that {\it `…a large swathe of the EoR window is corrupted by mutual coupling, and fringe-rate filtering seems to  only mildly alleviate the issue'}. Furthermore, {\it `simulated coupling is roughly an order-of-magnitude fainter than what is seen in the data accentuates the issue: mutual coupling poses a serious threat to a densely packed array’s ability to detect the cosmological 21-cm signal'}.

For the SKA, \citet{2016PASA...33...19T} and \citet{2017MNRAS.470..455T} considered the effect of bandpass smoothness and array configuration of the older SKALA2 and SKALA3 antenna designs on EoR statistical experiments. Unfortunately, antenna interactions, which are a crucial factor in chromatic behaviour, were not included. 
More recent work has focussed on electromagnetic modelling of the SKALA4.1 antennas for different SKA-Low station configurations, resulting in the availability of simulated scattering matrices  (which describe the complex interaction between antenna pairs) and embedded element patterns (which describe the beam pattern of individual antenna elements), inclusive of mutual interaction effects. These simulations, which use the HARP, Galileo and FEKO modelling tools \citep{2018ITAP...66.1805B,2022JATIS...8a1017B}, have been the basis for guiding the final design of SKA-Low antennas and stations. Nevertheless, the simulations are typically only available at discrete frequencies, making it difficult to judge the effect of chromaticity on spectral-line performance. 

This paper explores the accuracy of parameterisation of publicly available scattering matrices for different SKA-Low station configurations \citep{2022JATIS...8a1017B}, and develops delay-space methods to rapidly predict the point-source chromaticity for arbitrary SKA-Low antenna configurations, sky location and central frequencies.

Section~\ref{sec:method} of this paper introduces the impulse response methodology for assessing the spectral characteristics of radio telescopes. Section~3 discusses parameterisation of the complex scattering matrix and its application to SKA-low stations, including prototypes for which configuration data is available. Section~4 presents an analysis of SKA-low spectral response as a function of position on the sky. Section~5 presents a test-case simulation of a Cosmic Dawn experiment with an SKA-low station using different observing techniques; Section~6 introduces the effect of mutual coupling on station beam patterns; Section~7 further expands on issues relating to chromaticity, whilst Section~8 summarises the findings of this paper.

\begin{figure*}
    \centering
    \includegraphics[width=1.0\textwidth]{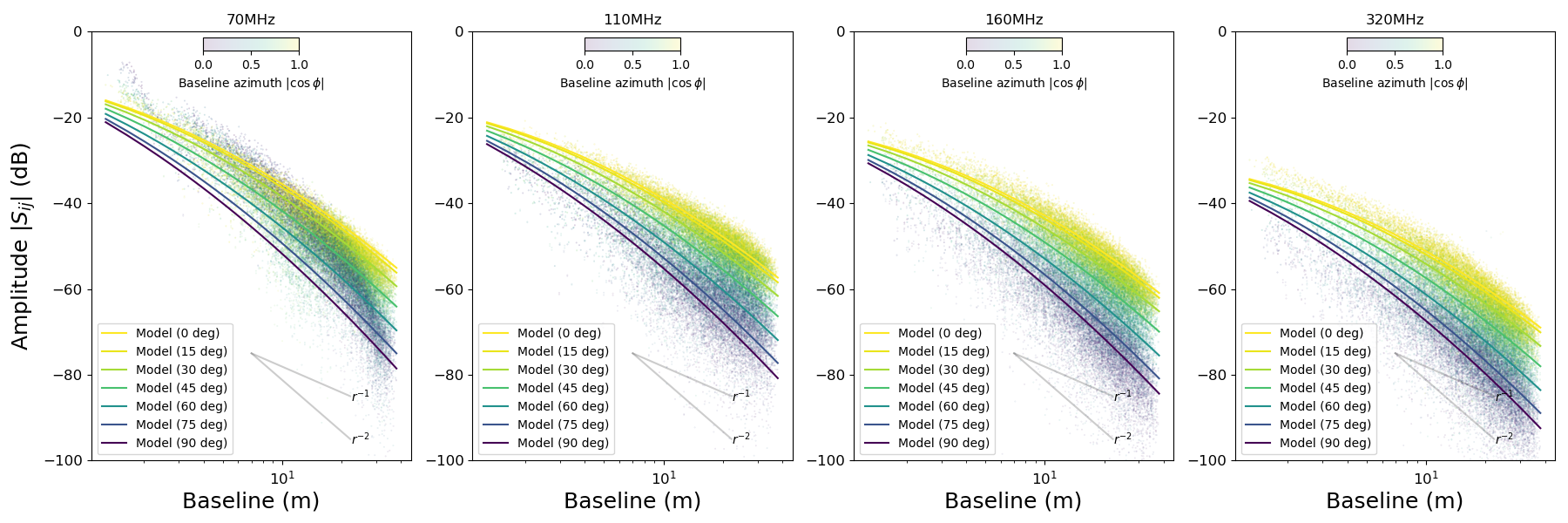}
    \caption{A multi-parameter fit to the multi-frequency AAVS2/SKALA4.1 $XX$ scattering matrix of \citet{2022JATIS...8a1017B}. The scattering amplitude falls steeply with baseline, and is sensitive to baseline azimuth.}
    \label{fig:scatter-amp}
\end{figure*}

\section{Method}
\label{sec:method}
A linear time-invariant (LTI) system can be characterised by its response to a delta function $\delta(\tau)$. The output $R(\tau)$ of an LTI system is the convolution of the input with the impulse response function (IRF) $g(\tau)$ of the system:

\begin{equation}
R(\tau) = g(\tau) * \delta(\tau) = \int_{-\infty}^{\infty} g(\tau - t) \delta(t) dt.
\label{eq:acf}
\end{equation}
The Weiner-Khinchen theorem applied to stationary random processes states that the complex frequency response of the system $A(\nu)$ is the Fourier transform of the IRF:

\begin{equation}
A(\nu) = \int_{-\infty}^{\infty} R(\tau) e^{-2\pi i \nu t} d\tau = \int_{-\infty}^{\infty} g(\tau) e^{-2\pi i \nu t} d\tau.
\label{eq:wk}
\end{equation}
For the purposes of assessing the response of an LTI system at high frequency resolution, the IRF method is a relatively straightforward and computationally efficient. It is used in time-domain reflectometers or transmissometers to measure the characteristics of electronic devices or the electromagnetic characteristics of physical objects. The IRF method is closely related to auto- and cross-correlation analysis of radio telescope signals, where Equation~\ref{eq:wk} is used to compute total- and cross-power spectra. 

This method is applicable to processes where there is no dependence between delay and frequency, or at least that any variance is small in the frequency range being investigated. Of course, in practical IRF measurements, it is important not to introduce non-linearities. However, this is not relevant here, where IRFs are used as a modelling tool only.


The IRF method allows the frequency response of radio telescopes to be predicted. This is particularly true for  antenna structures such as large reflectors such as 
the Murriyang (Parkes) telescope where `standing waves' can limit the spectral dynamic range of observations \citep{2017PASA...34...51R}.
Radiation from the telescope feed, either by leakage from the receiver or by reflection from the feed, is returned to the parabolic surface. Most of the reflected radiation will transmit away from the antenna. However, due to diffraction, a small portion will return to the focus and create an interference pattern. This is analogous to two-element interferometry, where the fixed delay difference between two antennas creates a fringe pattern in delay space as the baseline geometry changes, but also a fringe pattern in frequency space owing to the linear change of phase with frequency when the delay is constant.

Whilst the power levels of the returned radiation are small, Equation~\ref{eq:acf} refers to the electric field at the feed, or voltages in the transducer. This considerably magnifies the impact of the diffracted/reflected radiation. As an example, if we consider a single reflection with a total time delay $\tau_0$, Equation~\ref{eq:wk} tells us that the resultant power spectrum will be the product of a sinusoidal function (the Fourier Transform of a delayed delta function) and a flat white noise spectrum (the Fourier Transform of an undelayed delta function). The frequency of the sinusoid $\Delta\nu_0$ is

\begin{equation}
\Delta\nu_0 = \frac{1}{\tau_0}.
\label{eq:nu}
\end{equation}
The amplitude (i.e.\ $0.5\times$ the peak-to-peak) of the sinusoid $A_1$, relative to the power in the undelayed signal $A_0$, is given by

\begin{equation}
\frac{A_1}{A_0} \approx 2\gamma,
\label{eq:gamma}
\end{equation}
where $\gamma$ is the voltage ratio of the delayed and undelayed signals (assumed $\ll 1)$. Thus, a ripple amplitude of 1\%, relative to an underlying white noise power spectrum requires a scattered power of only 0.0025\%, or $-46$dB. Reflectometry measurements, even of large reflector antennas  \citep{Thomas1998},  indicate that such values are typical.

\section{Scattering}

Aperture arrays consist of multiple elements whose output is combined into a focus electronically rather than via a large reflector. This results in huge cost savings at low frequencies where large collecting areas are often required. It also results in potential benefits, such as wide fields of view. Examples already mentioned include MWA, LOFAR and the SKA low-frequency component, SKA-Low. 

The properties of such arrays are usually hard to model due to their complexity and large numbers of unknowns. Using the FEKO\footnote{Altair Engineering Inc. https://www.altair.com/feko.} and Galileo\footnote{Ingegneria Dei Sistemi S.p.A. https://www.idscorporation.com/pf/galileo-suite} electromagnetic analysis packages, \cite{2022JATIS...8a1017B} summarise the difficulty of electromagnetic analysis of the SKA Aperture Array Verification Station version 2 (AAVS2). Due to the large number of antenna unknowns ($10^4$), array unknowns ($10^6$) and slow convergence, their analysis was only been conducted at 14 discrete frequencies across the SKA-Low band (50--350 MHz). This is insufficient to judge the impact of design considerations on the ability of the SKA to detect spectral signatures associated with the Cosmic Dawn, the Epoch of Reionisation, redshifted 21-cm absorbers or post-EOR intensity mapping.

However, if frequency-dependent terms in the scattering matrix of the antenna elements change monotonically between analysis frequencies, it is possible to take sophisticated electromagnetic models and apply the IRF method to derive a model for the spectral response at better resolution, with less computational requirement.

\begin{table}
    \centering
    \begin{tabular}{lccccc}
    \hline
    Array   & $n$ & \multicolumn{4}{c}{Median Absolute Deviation (MAD)}   \\
            &       & $XX$ &  $YY$ & $XY$ & $YX$    \\
    \hline
    AAVS2   & 848,640 & 2.47dB & 2.50dB & 3.80dB & 3.80dB \\ 
    AAVS3   & 6,005,760 & 3.10dB & 3.10dB & 4.23dB & 4.23dB \\
    \hline
   
    \end{tabular}
    \caption{The accuracy of the multi-parameter fits to the AAVS2 $XX$ scattering matrices for frequencies greater than 50 MHz when applied to other arrays and polarisations for which matrices are available \citep{2022JATIS...8a1017B}. $n$ is the number of non-diagonal elements in the scattering matrices to which the fit is applied.}
    \label{tab:scatter-fit}
\end{table}

\begin{table}
    \centering
    \begin{tabular}{lccccc}
    \hline
    Array   & $n$ & \multicolumn{4}{c}{Median Absolute Deviation (MAD)}   \\
            &       & $XX$ &  $YY$ & $XY$ & $YX$    \\
    \hline
    AAVS2   &  848,640   & $23.3^{\circ}$ & $23.7^{\circ}$ & $29.2^{\circ}$ & $29.2^{\circ}$ \\
    AAVS3   &  6,005,760 & $25.3^{\circ}$ & $26.0^{\circ}$ & $31.4^{\circ}$ & $31.4^{\circ}$ \\
    \hline
   
    \end{tabular}
    \caption{The median absolute deviation between the scattering phase model and the scattering matrix phases, for frequencies greater than 50 MHz (see four example frequencies in Figure~\ref{fig:scatter-ph}). $n$ is the number of non-diagonal elements in the scattering matrices.}
    \label{tab:scatter-phase}
\end{table}

\begin{figure*}
    \centering
    \includegraphics[width=1.0\textwidth]{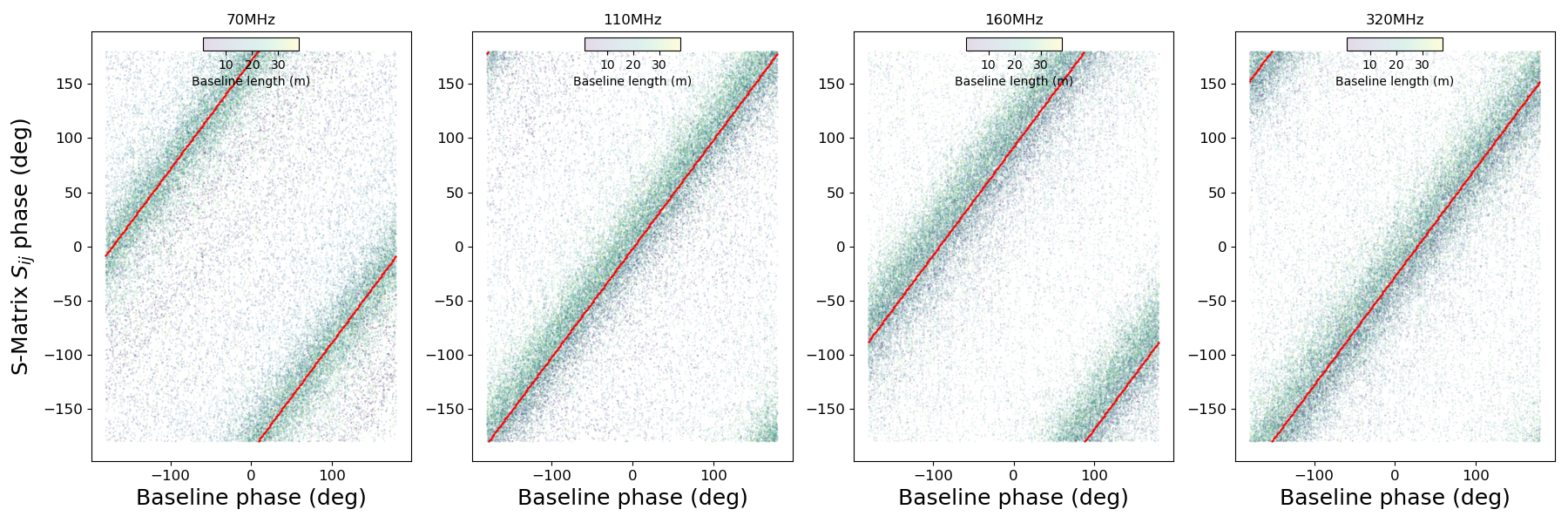}
    \caption{Geometric phase (calculated from AAVS2 array geometry and frequency) versus S-parameter phase estimated using {\it FEKO} by \citet{2022JATIS...8a1017B}. The systematic offsets from the origin of the S-parameter phases are mostly due to antenna characteristics rather than mutual coupling. The red line was calculated from a fit in frequency space to the unwound phase offsets from the denser AAVS3 S-parameters (i.e. not a fit to the AAVS2 data shown).}
    \label{fig:scatter-ph}
\end{figure*}

\subsection{Amplitudes}
\label{sec:scatter-amp}

We have derived approximations to the amplitudes of the FEKO scattering matrices $|S_{ij}|$ computed for SKALA4.1 antennas in the AAVS2 configuration by \citet{2022JATIS...8a1017B}. Amplitudes for all $256^2$ baselines in the $XX$ scattering matrix are shown in Figure~\ref{fig:scatter-amp} (excluding the 256 autocorrelations along the diagonal) for four example frequencies. The major sources of variance in the scattering matrix are frequency, baseline length and baseline azimuth. The antenna voltage transmission efficiencies are also relevant in explaining the low coupling at the lowest frequencies -- these were fixed according to the estimates of \citet{2022JATIS...8a1017B}. A multi-parameter fit with six free parameters (scipy {\tt least\_squares} with a soft L1 loss function) to all AAVS2 matrices ($256^2$ elements $\times 14$ frequencies, minus auto-correlations) results in the model lines shown in Figure~\ref{fig:scatter-amp}. As listed in Table~\ref{tab:scatter-fit}, the AAVS2 fit has a median absolute deviation of 2.5dB, with most of the deviation occurring at short baselines where the effects of near-field mutual coupling are most variable. 

The same fit parameters\footnote{The projection of the electric field is $\cos \phi$ for $XX$ matrices, $\sin \phi$ for $YY$  matrices and $\sin  2\phi$ for $XY$ matrices, where $\phi$ is the baseline azimuth.}, if applied to the AAVS2 $YY$ matrix or the AAVS3 FEKO scattering matrices (each with $256^2$ elements $\times 93$ frequencies, minus auto-correlations), result in a similar median absolute deviation ($\sim 3$dB, see Table~\ref{tab:scatter-fit}). This is a small fraction of the 100dB, or so, full dynamic range of the scattering amplitudes apparent in Figure~\ref{fig:scatter-amp}. The same fit parameters, with a 5dB offset to reflect the overall lower cross-coupling between orthogonal polarisations, are also useful for the $XY$ matrices, though with a higher MAD of 4dB. However, a free fit with the same functional form again reduces the cross-polar MAD to $\sim 3$dB. 

Together, these results strongly suggest that amplitude parameterisation of the scattering matrices is feasible and, moreover, reasonably independent of array configuration. These results are supported by the strong similarity in the antenna element self-interactions within and between difference array configurations (the diagonals of the scattering matrices)

\subsection {Delays and phases}

Figure~\ref{fig:scatter-ph} shows the high degree of correlation of geometric phases, calculated from baseline lengths and frequency, and the negated scattering matrix phase\footnote{This accounts for the difference between engineering ($e^{j\omega t- k.r}$) and physics ($e^{k\cdot r-i\omega t}$) conventions.}, again demonstrating the ease of separating antenna properties from array properties. However, common to all antenna elements and array configurations, there are fixed but frequency-dependent phase offsets such that the mean phases do not pass through the origin in Figure~\ref{fig:scatter-ph}. Some of this is due to the frequency-dependent focus of the SKALA4.1 elements (see Section~\ref{sec:ground}), but much will arise from phase offsets in the internal line feed in the antenna. 

Nevertheless, after subtracting geometric phase, the frequency-phase response across non-diagonal elements of the scattering matrix can be unwound, and is the same for all baselines and array configurations, again indicating that antenna and array responses are separable. 
 The scattering matrices reveal a phase ramp of $-24$ deg/MHz at a frequency of 90 MHz once geometric phase is accounted for, rising to $-10$ deg/MHz at 350 MHz. The resultant phase acceleration is 0.05 deg/MHz$^2$. 
 As long as channel bandwidths are sub-MHz, the consequences of such a large phase ramp ($-84.4$ rad across the band)\footnote{\citet{2017MNRAS.470..455T} report a much smaller value of $-10$ rad for the SKALA3 antennas} are not important for interferometry between similar antennas. But, if genuine, this may be problematic for interferometry with non-SKALA antennas.

\section{Spectral modelling}
\label{sec:spectral}

\begin{figure*}
    \centering
    \includegraphics[width=0.48\textwidth]{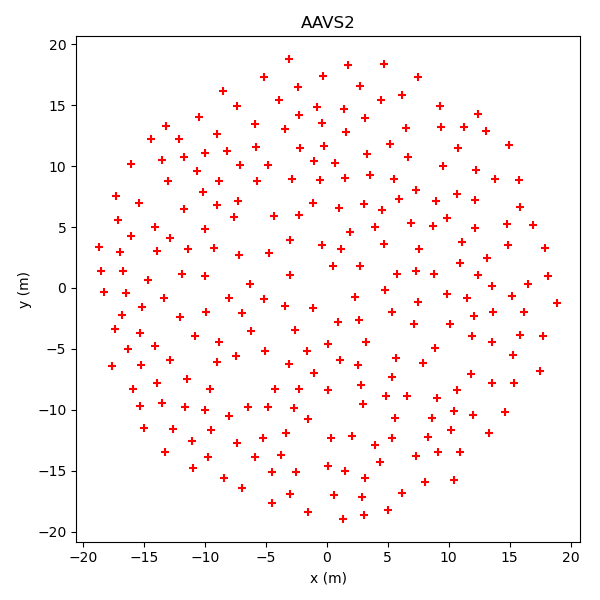} 
    \includegraphics[width=0.48\textwidth]{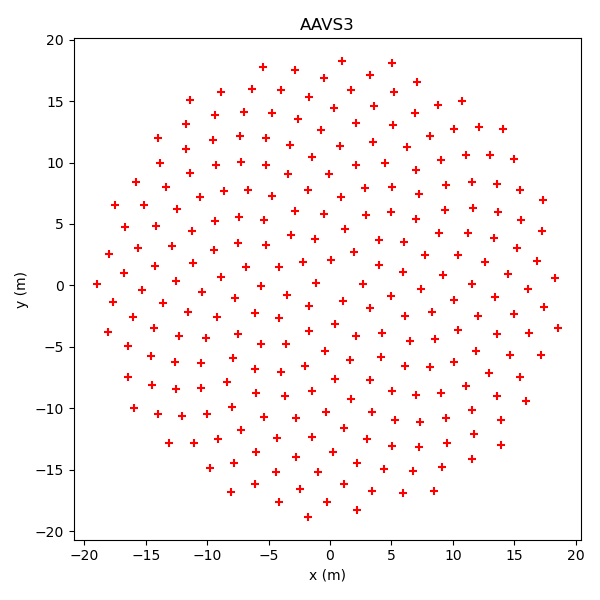}
    \includegraphics[width=0.48\textwidth]{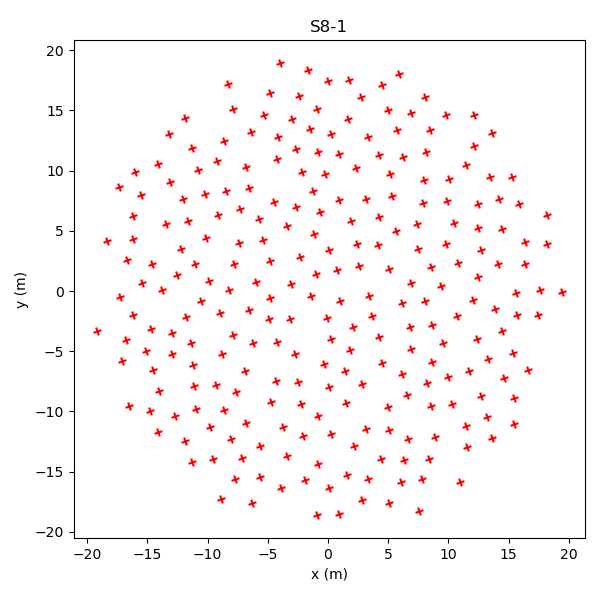} 
    \includegraphics[width=0.48\textwidth]{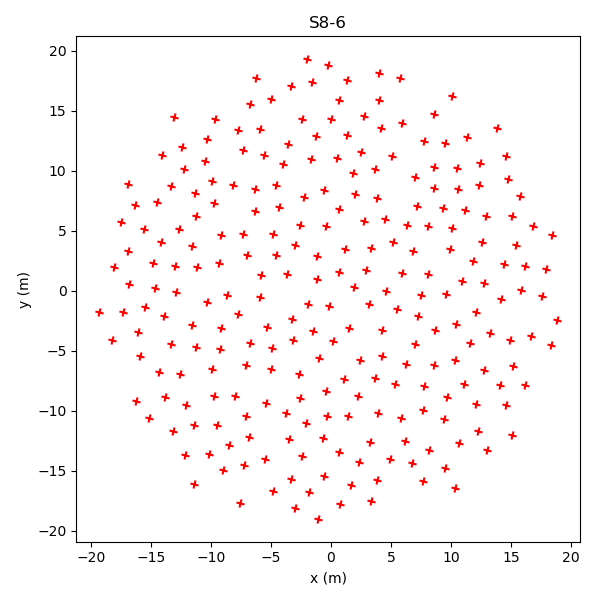}
    \caption{SKA-Low station configurations for the prototype stations AAVS2 and AAVS3; and the first deployed stations, S8-1 and S8-6. S8-1 and S8-6 are rotated in azimuth by 251.3deg and 193.6deg, respectively.}
    \label{fig:antenna_layout}
\end{figure*}

\begin{figure*}
    \centering
    \includegraphics[width=1.0\textwidth]{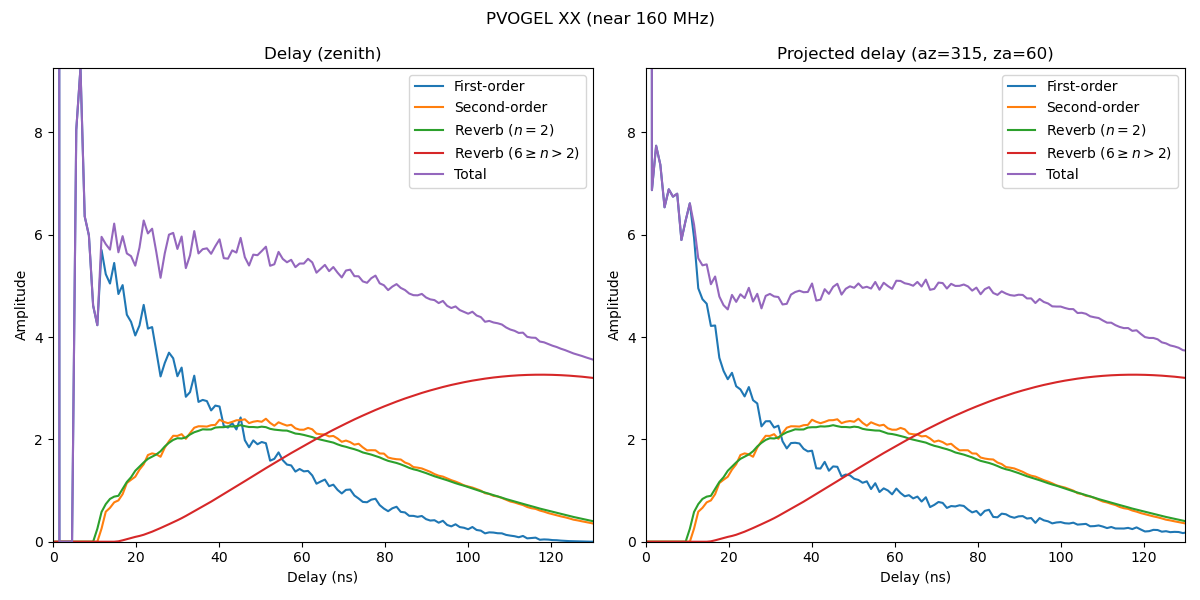}
    \caption{The computed delay spectrum for an SKA-Low S8-1 station with no ground reflections when pointing at: (left) the zenith; and (right) azimuth 315 deg, zenith angle 60 deg. The amplitude at zero delay is normalised at 256 (the number of antenna elements). Scattered amplitudes are typically 1.5--2.5\% over the delay range shown (which approximately corresponds to the 38-m station diameter). Contributions to first-order ($n=1$), second-order ($n=2$) and multiple reflections ($6\geq n>2)$ are shown separately. The orange lines are a brute force calculation involving dual reflections. The green and red reverberation lines are computed using the convolution theorem (a polynomial series of Fourier transforms).}
    \label{fig:delay-spectrum}
\end{figure*}

\subsection{2D Array Model}
To examine the effect of station array configuration in a given pointing direction, it is assumed that the SKA-Low observing system is phased up so that all primary signals arrive simultaneously at the beamformer. A fraction of the electric field (18--100\%, depending on frequency) is reflected from each antenna and a fraction of that (closely determined by the scattering matrix) will be received at all other antennas, resulting in interference. In fact, due to the high reflection coefficients of the SKALA4.1 antennas at some frequencies, multiple reflections are possible. These reverberations can be tracked by a polynomial series of Fourier transforms of the first-bounce delay spectrum. At most frequencies, only the first reflection contributes to the final power spectrum. At frequencies below 100 MHz, there are a large number of reflections, resulting in finer bandpass structure. 

For the purposes of better separating antenna and array contributions, we begin by examining the electric field in the aperture plane rather than at the antenna output ports (to which the scattering matrices refer). This removes the highly-variable contribution of the frequency and angular response of the antenna elements. However, it does mean that the amplitudes of the scattering matrices will be slightly underestimated. Therefore, in amplitude, we divide by the frequency-dependent voltage transmission coefficient to better represent the increased scattered radiation in the aperture plane relative to the output ports (above 90MHz, this correction is very minor). In the 2D array case, we assume that the only geometric delays are due to the array configuration.

In this paper, where we will only consider Stokes $I$ sky models, we ignore cross-polar coupling. However, dual cross-polar reflections will create interference at a level only 10dB below co-polar dual reflections, so will be more important than third-order co-polar reflections. Since these will only be important at the lowest frequencies, we ignore these for the current purposes.

We consider four SKA-Low station configurations: the two prototype stations AAVS2 and AAVS3; and the first deployed stations, S8-1 and S8-6. AAVS2 and the deployed S8 stations have antenna configurations which are pseudo-random. The AAVS3 configuration is a Vogel pattern \citep{2023ursi.confE..22D}. AAVS2 and AAVS3 have elements whose principle $X$ and $Y$ axes are aligned with geographic E-W and N-S. S8-1 and S8-6 are the deployed SKA-Low station configurations, but rotated  by 251.3deg and 193.6deg, respectively. The station layouts are shown in Figure~\ref{fig:antenna_layout}.

Using the S8-1 station as an example, the delay spectra can be computed using the fits to the scattering matrix in Section~\ref{sec:scatter-amp} and the purely 2D geometric delay terms. An example is shown in Figure~\ref{fig:delay-spectrum} at 160 MHz. At the zenith, the delayed signal peaks at 3.6\% of the direct signal at 7 ns, which unsurprisingly corresponds to the nearest antenna distances.

At 160 MHz, more than half the signal which is delayed by more than 40 ns arises from double reflections, where the incident signal has reflected off two antenna elements before interfering with the primary wavefront. At frequencies below 90 MHz, where the SKALA4.1 reflection coefficient becomes very high, such multiple reflections (or reverberations) dominate, and will give rise to rapidly changing spectral structure at the MHz scale. 

\begin{figure*}
    \centering
    \includegraphics[width=1.0\textwidth]{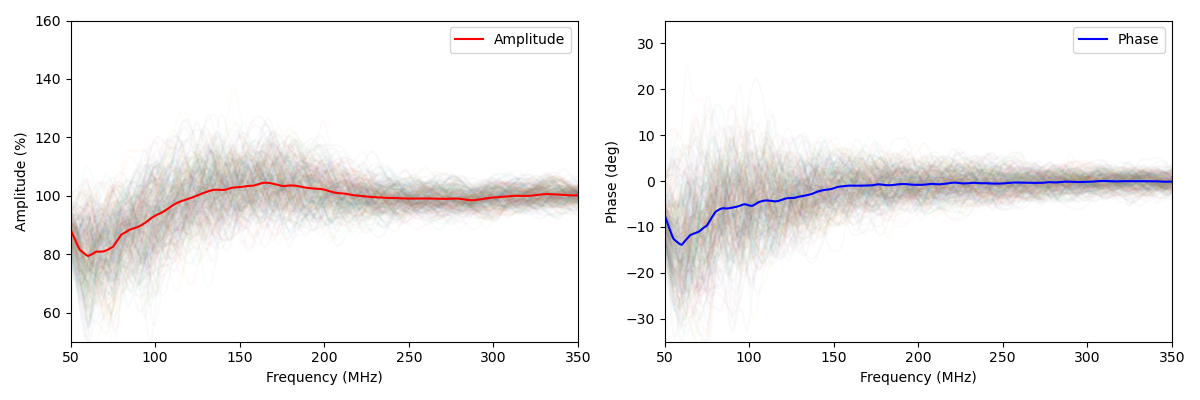}
    \includegraphics[width=1.0\textwidth]{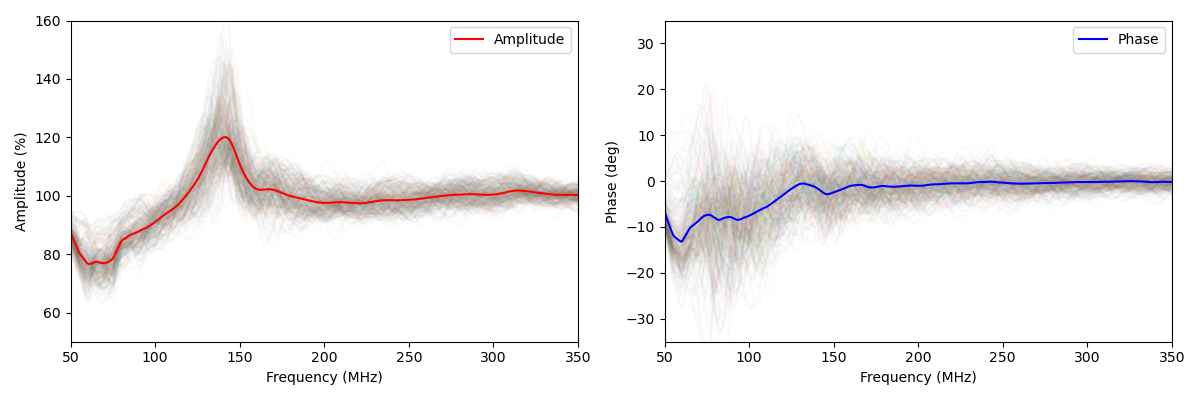}
    \includegraphics[width=1.0\textwidth]{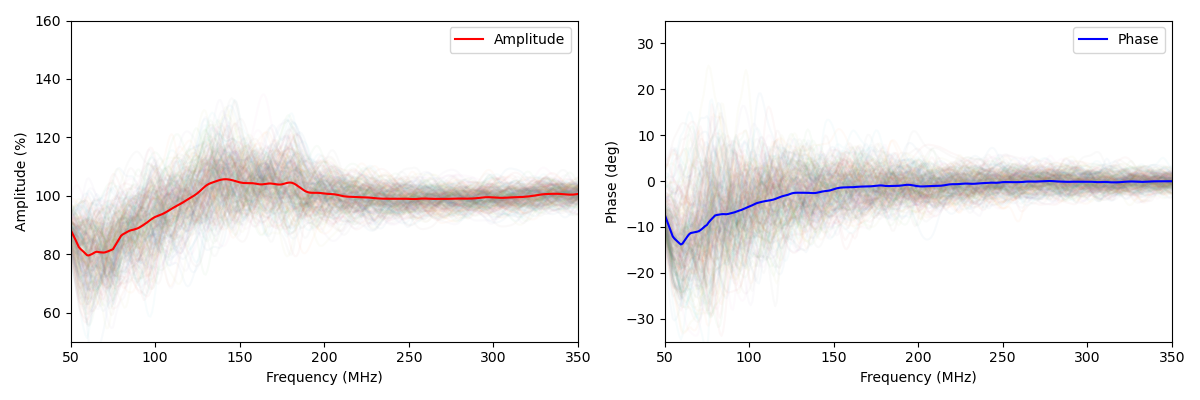}
    \caption{The predicted (left column) amplitude and (right column) phase power spectra for (top to bottom) the SKA-Low AAVS2, AAVS3 and S8-1 configuration when pointing at the zenith. The S8-1 and S8-6 zenith responses are identical. The mean phased station responses are shown by the solid solid lines (red for amplitude, blue for phase); the aperture fields associated with the 256 individual antennas are shown by the fainter lines. This calculation excludes antenna delays and ground reflections.}
    \label{fig:power-spectrum}
\end{figure*}

\begin{figure*}
    \centering
    \includegraphics[width=1.0\textwidth]{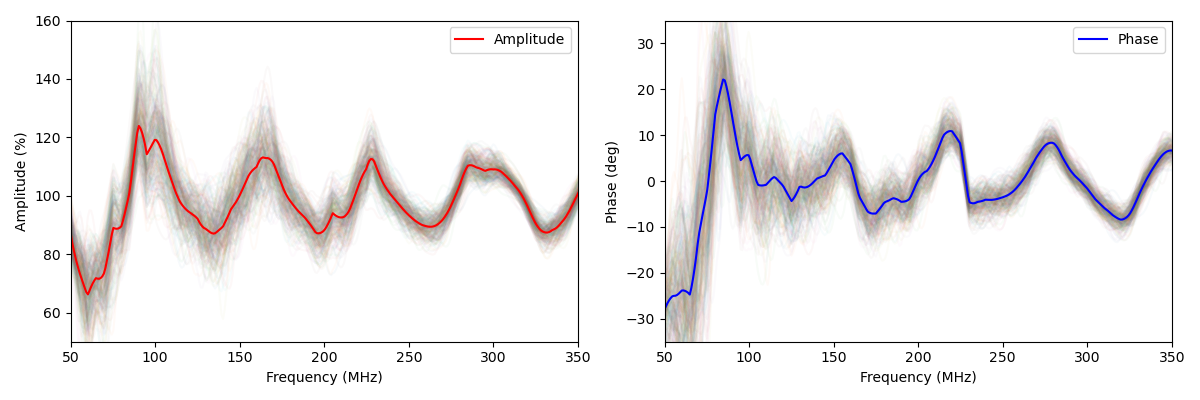}
    \includegraphics[width=1.0\textwidth]{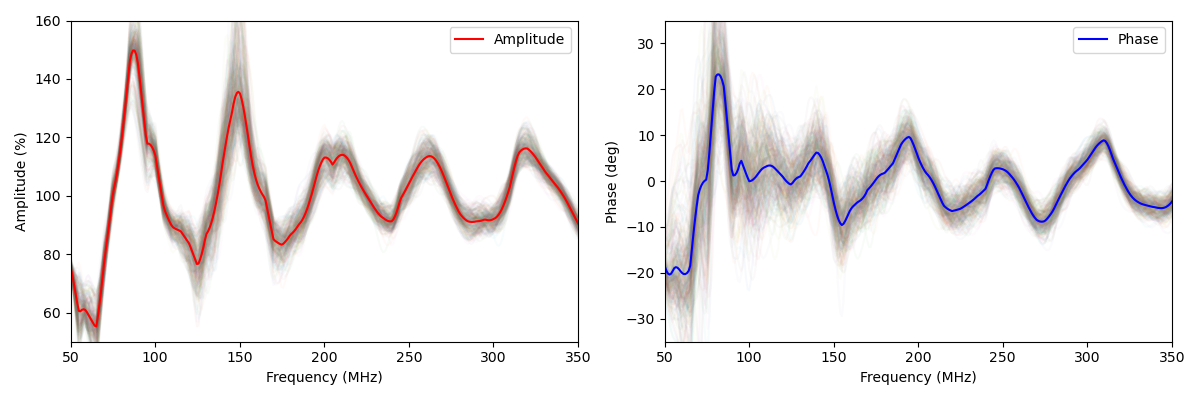}
    \includegraphics[width=1.0\textwidth]{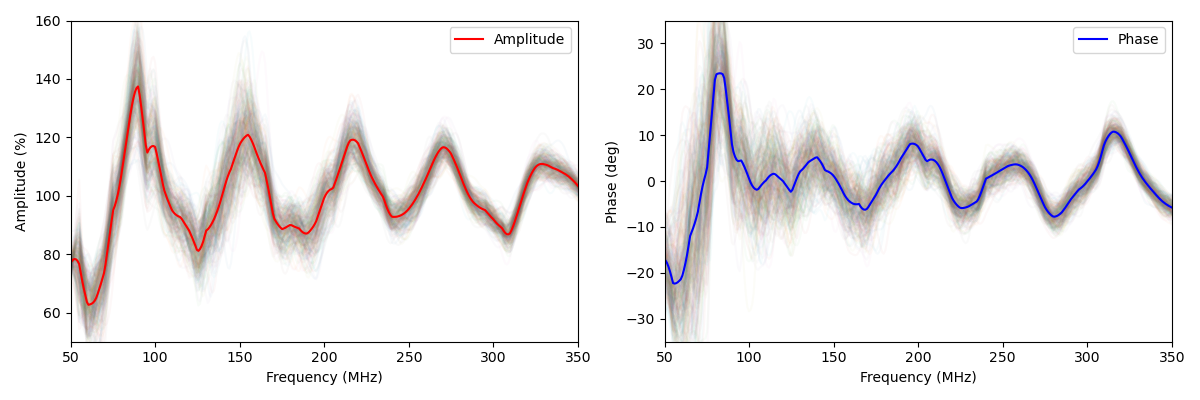}
    \caption{The predicted (left column) amplitude and (right column) phase power spectra for (top to bottom) the SKA-Low AAVS2, AAVS3 and S8-1 configuration when pointing at the zenith. The S8-1 and S8-6 zenith responses are identical. The mean phased station responses are shown by the solid solid lines (red for amplitude, blue for phase); the aperture fields associated with the 256 individual antennas are shown by the fainter lines. This calculation includes ground reflections.}
    \label{fig:power-spectrum-ground}
\end{figure*}

As discussed in Section~\ref{sec:method}, the delay spectrum shown in Figure~\ref{fig:delay-spectrum} can be Fourier transformed to give power spectrum over that part of frequency bandwidth where the amplitude of the scattering matrix is reasonably constant. In our analysis, we typically use a window of $\sim 5$ MHz, corresponding to a bandwidth/frequency ratio varying from 1.4\% to 10\% over the SKA-Low frequency range. Assuming that most of the aperture-plane power spectrum structure arises from phase differences resulting from the geometric delays, this approach is very robust. So-called XF spectrometers, formerly popular in radio small-$n$ radio interferometry, would typically operate over larger $\sim10$\% bandwidths. 

Example zenith power and phase spectra of deployed SKA-Low stations and prototypes are shown in Figure~\ref{fig:power-spectrum}. As these are calculated in the aperture plane of each station, the spectra shown do not reflect the gross frequency response \citep{2020IOJAP...1..253B} or sky-dependence (see Section~\ref{sec:Mock}) of the SKALA4.1 elements. 

The (vector) average of the S8-1 station beam (bottom of Figure~\ref{fig:power-spectrum}) deviates from unity by up to 5--10\% and 5--10 deg, but the fields above individual antennas can vary by 30\% and 30 deg, depending on where they reside within the array. The regular array configuration of AAVS3 (middle plot) has power and phase frequency structure that is typically of factor of two worse as a result of baseline periodicity, although this improves greatly for non-zenith pointings. 

\subsection{Incorporating Ground Reflections}
\label{sec:ground}

The 3D nature of the SKALA4.1 antenna further complicates the scattering geometry of SKA-Low arrays and must be accounted for in calculating the final aperture array response. 
According to \citet{2022JATIS...8a1023B}, the metallic ground screen creates a reflected field component which adds to the field received by the SKALA4.1 antennas. This is in addition to the field directly reflected by other antennas, as discussed above. The ground screen creates additional delays, especially at the highest frequencies where the receptors are furthest from the screen. \citet{2022JATIS...8a1023B} also detail the difficulties faced in accurately calculating this component. In this study, we use a simple interpolation model to estimate the frequency-dependent dipole height above the ground plane and factor in the extra delay from the ground plane back to the dipole of the antenna receiving the scattered radiation.

The ground reflections modify the scattering geometry between antenna pairs, between multiple antennas, and between an antenna element and itself. In particular, some of the radiation not absorbed by an antenna will scatter from the ground screen back into the antenna, creating interference, often known as a standing wave. The average SKALA4.1 voltage reflection coefficient is very high \citep[$\sim 0.2$ above 90MHz;][]{2022JATIS...8a1017B}. This self-interfering radiation will mainly arise from a dipole-shaped ground patch around each antenna. 

The predicted SKA-Low power and phase spectra for the same SKA-Low station configurations are shown in Figure~\ref{fig:power-spectrum-ground}. These calculations are also given in the aperture plane, but include the effects of array geometry, antenna geometry and ground-plane reflections. They all show strongly chromatic behaviour by way of a pseudo-sinusoidal $\sim60$ MHz periodicity due to a strong delay peak at around 11--16ns. As seen by comparison with Figure~\ref{fig:power-spectrum}, this is entirely due to ground-plane reflections, mainly self-interference. There is frequency-dependence in the additional ground-screen delay, but due to the extent of the ground-plane (we assume contributions to self-interference extend out to the minimum baseline length), the additional delay increases by only 40\% between frequencies 100 and 350 MHz. Figure~\ref{fig:power-spectrum-ground} shows that the station beam gains deviate from their average by 20--40\% in amplitude and 10--20 deg of phase, which is several times greater than the contributions from array configuration alone.

\section{Mock observations}
\label{sec:Mock}
Using the fact that a band-limited IRF technique can broadly reproduce the chromatic characteristics of SKA-Low stations efficiently at high spectral resolution and for arbitrary array/sub-array configurations, we demonstrate the use of the technique in conducting mock observations with different array configurations.

\subsection{Sky continuum model}

\begin{figure*}
    \centering
    \includegraphics[width=1.0\textwidth]{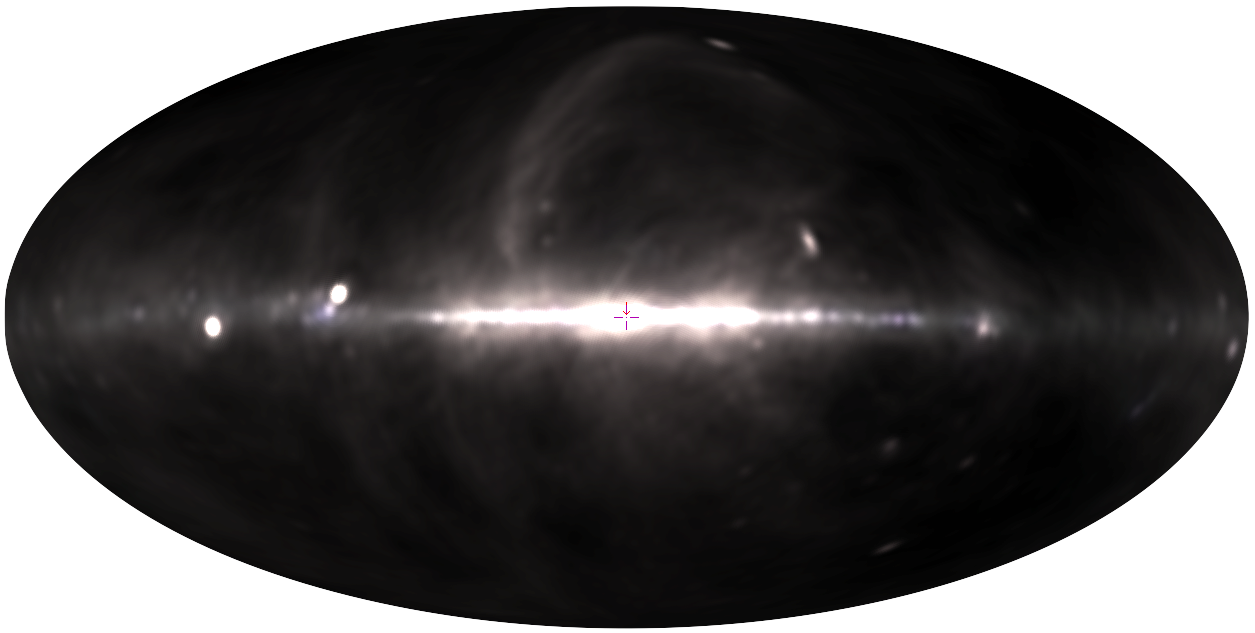}
    \caption{A three-colour RGB (150, 250 and 350 MHz) rendition of the Low Frequency Sky Model of \citet{2017MNRAS.469.4537D} in Galactic coordinates at an angular resolution of 3 deg. }
    \label{fig:LFSM}
\end{figure*}

We use the Low Frequency Sky Model (LFSM) created by \citet{2017MNRAS.469.4537D}. It conveniently covers the SKA-Low frequency range (50--350 MHz) and is based on a Principle Component Analysis of sky maps at 16 frequencies, including the LWA at 40-80 MHz \citep{2017MNRAS.469.4537D} and all-sky maps at 408 MHz \citep{1982A&AS...47....1H, 2015MNRAS.451.4311R}. We downloaded the data at a frequency resolution of 1 MHz and an angular resolution of 3.0 deg. Since the angular resolution of SKA-Low stations is similar at $\sim 150$ MHz, there was no compelling reason to combine with data at higher resolution, for example from the GLEAM survey \citep{2017MNRAS.464.1146H} which covers around 50\% of the SKA-Low frequency range. An RGB version of the sky model is shown in Figure~\ref{fig:LFSM} in Galactic coordinates.

\subsection{Cosmic Dawn model}
The expected `Cosmic Dawn' 21-cm signal has yet to be definitively detected \citep{2018Natur.555...67B, 2022NatAs...6..607S}, but it is a fairly robust prediction of standard cold dark matter cosmology and represents the moment at which radiation from the first stars begins to heat up the otherwise cold interstellar/intergalactic medium. Predicted redshifts lie within the range $10<z<25$ (55--130 MHz). The signal is expected to have an all-sky component with small angular fluctuations. The  frequency width ($10$s of MHz) and S/N ratio of the all-sky component is similar to that expected from EOR fluctuations. The expected higher amplitude of the all-sky Cosmic Dawn signal ($\sim0.1$ K in absorption) relative to the EOR signal is offset by the much higher Galactic and extragalactic foreground contamination at these lower frequencies. Interferometric detection of the Cosmic Dawn signal, via measurement of its power spectrum, is a key SKA science goal \citep{2015aska.confE...1K}. However, detection of the corresponding all-sky signal is not \citep[but see][]{2023JApA...44...24R}. We therefore use this example purely as a test case to judge station performance versus theoretical noise-free performance. Extension of the above results to cross-station applications, including imaging and power spectrum measurement, is a closely-related problem as the interferometer response is the product of the respective voltage gains, after accounting for any mutual cross-station coupling.

\begin{figure}
    \centering
    \includegraphics[width=1.0\textwidth]{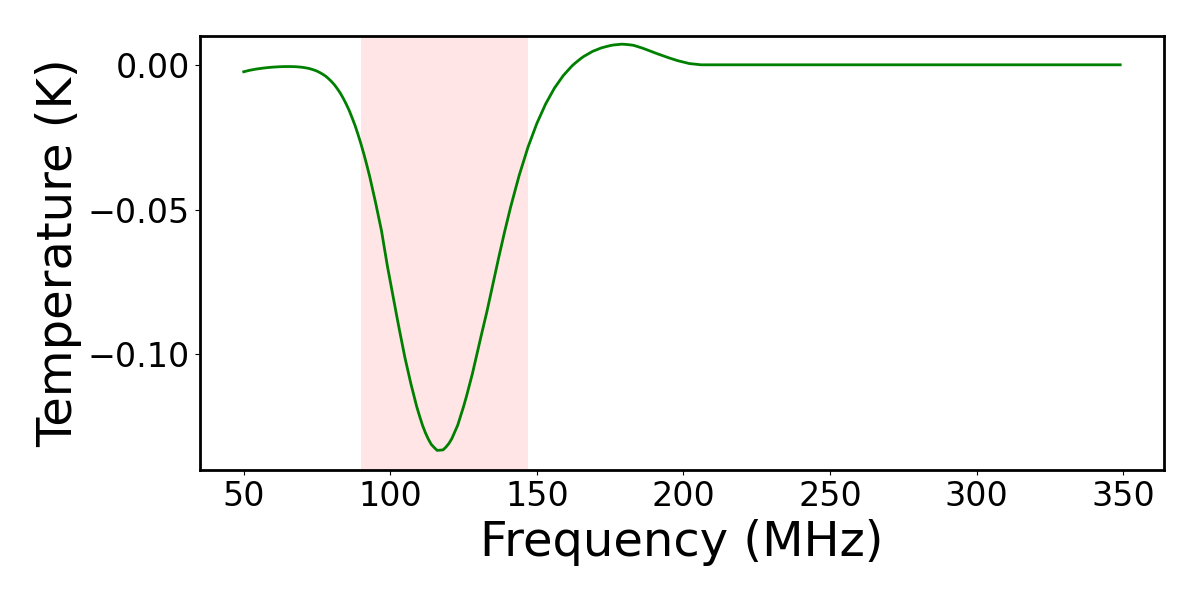}
    \caption{The redshifted 21-cm Cosmic Dawn signal predicted with 21CMFAST/21CMEMU using the parameter set in Table\ref{tab:21cmfast} and Planck 2018 cosmologyis shown by the solid green line. The width of the signal at 20\% of the peak absorption is shown with the light red shaded area.}
    \label{fig:21cmfast}
\end{figure}

We use 21CMFAST \citep{2011MNRAS.411..955M} and 21CMEMU \citep{2024MNRAS.527.9833B} to simulate Cosmic Dawn signals. For test purposes, we assumed a flat CDM Planck 2018 cosmology \citep{2020A&A...641A...6P} with ($\Omega_{\Lambda}$, $\Omega_{\rm m}$, $\Omega_{\rm b}$, $h$, $\sigma_8$, $n_s$) = (0.69, 0.31, 0.049, 0.68, 0.82, 0.97) and  the 21CMEMU parameter set shown in Table~\ref{tab:21cmfast}. This results in an all-sky absorption signal of $-0.133$ K centred at 117 MHz ($z=11$), as shown in Figure~\ref{fig:21cmfast}. The 50\% and 20\% widths of the absorption dip are approximately 43 and 57 MHz, respectively.

\begin{table}
    \centering
    \begin{tabular}{lcc}
    \hline
    Parameter      & Value  \\
                   &        \\
    \hline
Stellar-to-halo mass coefficient, $\log_{10} f_{*,10}$    & 0.58 \\
Stellar-to-halo mass power law, $\alpha_*$              & 0.65 \\
UV escape coefficient, $\log_{10} f_{\rm esc, 10}$ & 0.51 \\
UV escape power law, $\alpha_{\rm esc}$      & 0.85 \\
AGN turn-on mass, $\log_{10} M_{\rm turn}$ (M$_{\odot}$)   & 0.43 \\
SFR timescale, $t_*/H$                   & 0.51 \\
Normalised X-ray luminosity, $\log_{10} (L_{\rm X<2keV}/{\rm SFR})$ (erg\ s$^{-1}$ M$_{\odot}^{-1}$ yr) & 0.46 \\
X-ray escape threshold, $E_0$ (keV)         & 0.03 \\
X-ray power law, $\alpha_{\rm X}$        & 0.55  \\
    \hline
   
    \end{tabular}
    \caption{21CMFAST parameters \citep{2024MNRAS.527.9833B} used to calculate the Cosmic Dawn signal example shown in Figure~\ref{fig:21cmfast}.}
    \label{tab:21cmfast}
\end{table}

\subsection{Noise model}
Two types of noise are considered here:
\begin{enumerate}
    \item Thermal noise, or white noise, is calculated using the single-polarisation radiometer equation $\Delta T = T_{\rm rx}/\sqrt{\Delta \nu \Delta t}$, where $T_{\rm rx}$ is the sum of the receiver noise noise (assumed to be 40 K, including ground spillover, over the SKA-Low frequency range) and the sky background from LFSM, $\Delta \nu$ is the frequency resolution (Hz), and $\Delta t$ is the integration time (s). For Stokes $I$ ($=(XX+YY)/2$) measurements of unpolarised sources, $\Delta T$ is further reduced by $\sqrt{2}$.
    \item $1/f$ noise, or pink noise, is a common feature in analogue electronics and describes correlated variations that are typical in the gain and frequency response of amplifiers and other analogue components. SKA-Low system requirements for station beam and gain stability are $\sim 0.05$\% and $\sim 0.03$\% over a 600 s time interval, but unspecified frequency interval \citep{2017Caiazzo}. In practice, higher systematic LNA gain variations have been measured on prototype SKA station elements \citep[e.g.\ 2\% in power for every degree of temperature change;][]{2019Waterson}.
\end{enumerate}
    
Without further calibration, $1/f$ noise is therefore likely to dominate white noise for any bandwidth--time product in excess of $10^6$ (e.g. 1s/10kHz or 0.01s/1MHz) when in station mode. For interferometry between stations, requirements on $1/f$ noise are less stringent as gain variations apply only to the correlated signal. Furthermore, better calibration options are available. 
    
For the current purposes, we normalise the rms $1/f$ (strictly, $1/\sqrt{f_t f_{\nu}}$) gain noise in the 2D time-frequency plane to 0.09\% over each whole day and over the whole SKA-Low frequency range of 300 MHz.

\begin{figure*}
    \centering
    \includegraphics[width=0.48\textwidth]{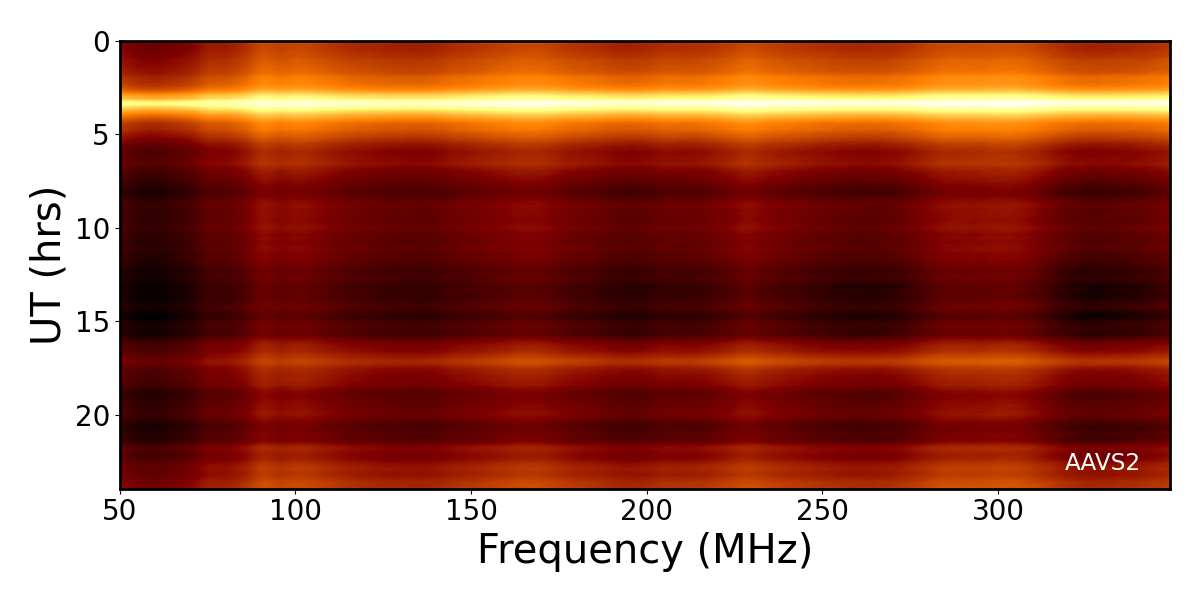}
    \includegraphics[width=0.48\textwidth]{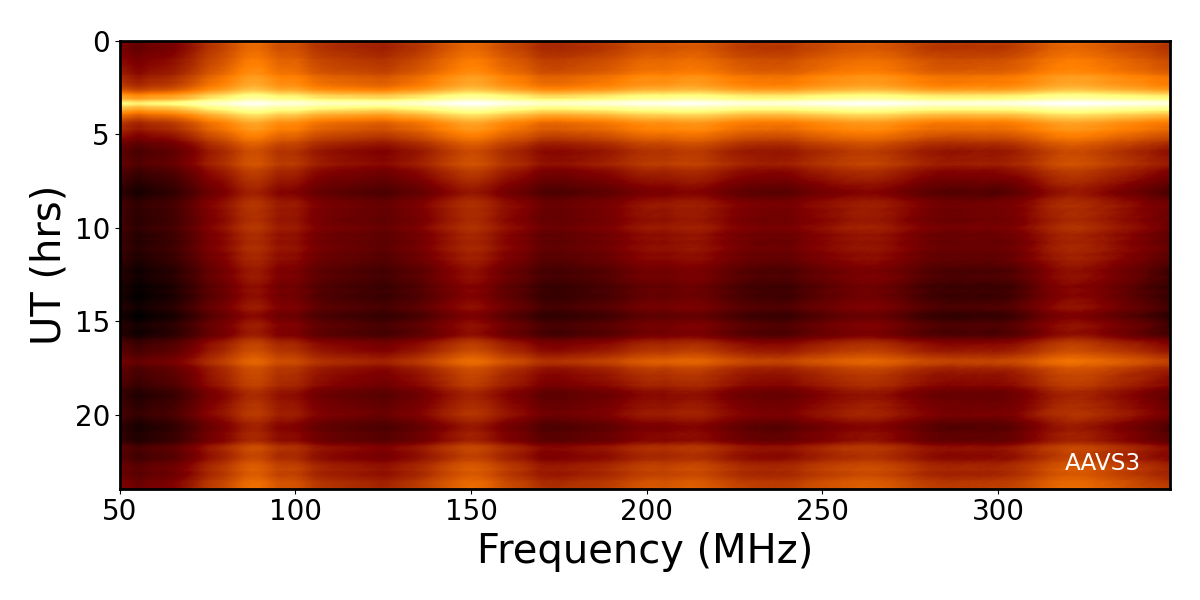}
    \includegraphics[width=0.48\textwidth]{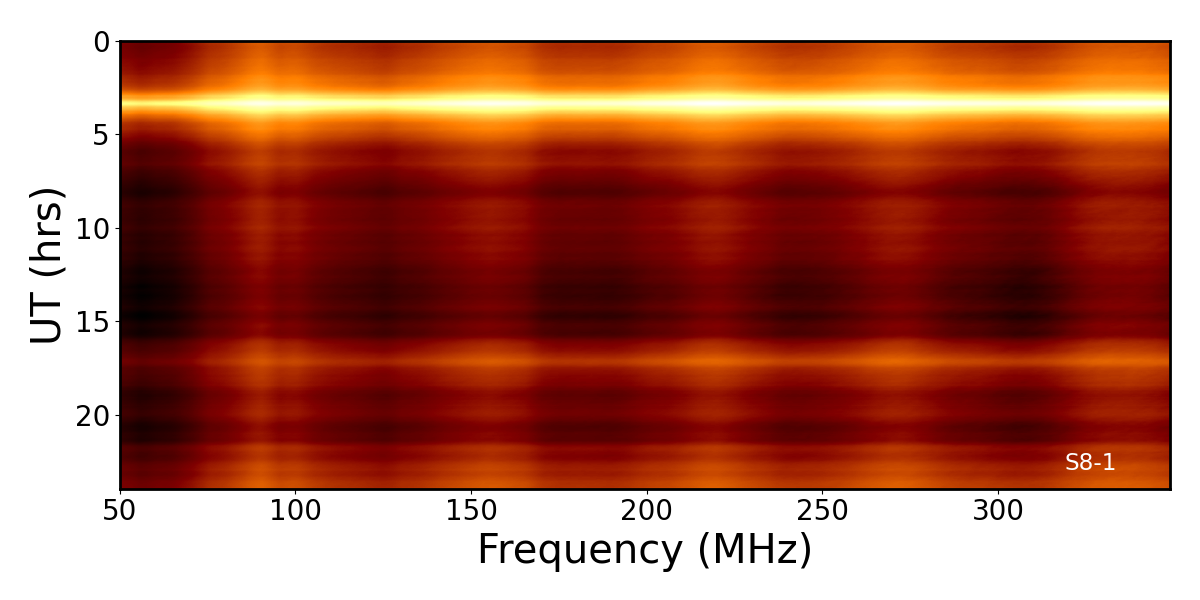}
    \includegraphics[width=0.48\textwidth]{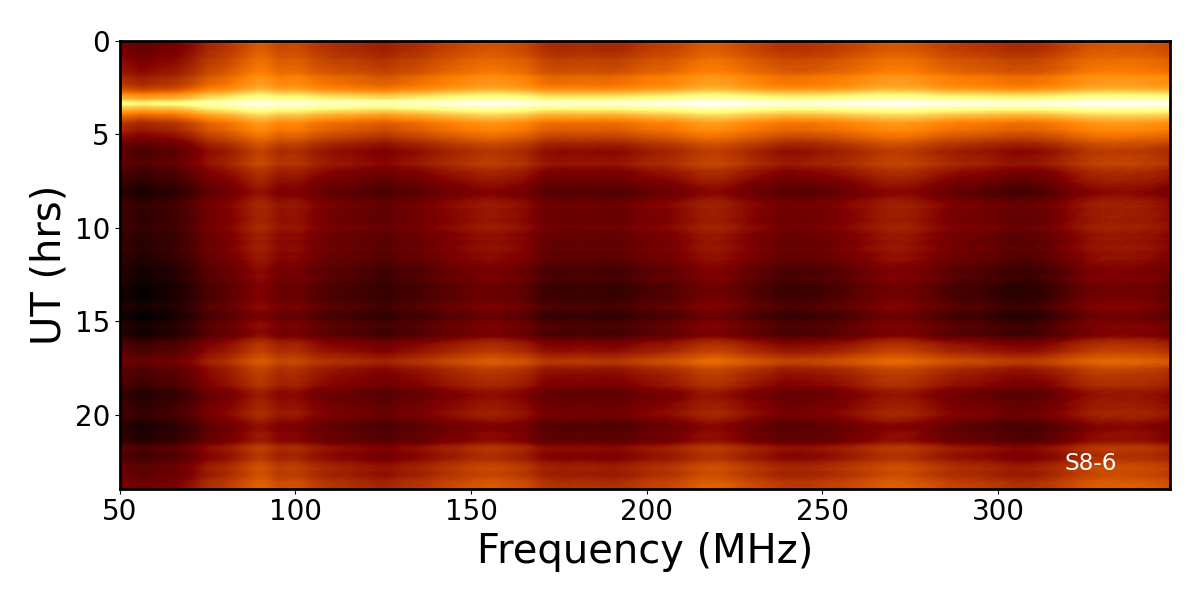}
    \caption{Simulated Stokes $I$ drift scans (frequency in MHz vs time in hrs) for four SKA-Low station configurations. The Low Frequency Sky Model of \citet{2017MNRAS.469.4537D} has been multiplied by the spectral gain model for the following configurations: AAVS2 (top left); AAVS3 (top right); S8-1 (bottom left); and S8-6 (bottom right). For clarity, the spectra have been de-reddened assuming a uniform spectral index of $-2.7$ and a normalisation frequency of 160 MHz. An artificial cosmic dawn signal has been added, but is too weak to be seen here. Noise has been added as described in the text. Horizontal lines are radio sources; vertical lines are the modelled gain fluctuations in the aperture plane as a result of intra-station antenna interactions. The simulation is based on the sky passing through the zenith on 1 January 2025. The de-reddened intensity range is 130 K to 4500 K (logarithmic scale).}
    \label{fig:drift}
\end{figure*}

\begin{figure*}
    \centering
    \includegraphics[width=0.48\textwidth]{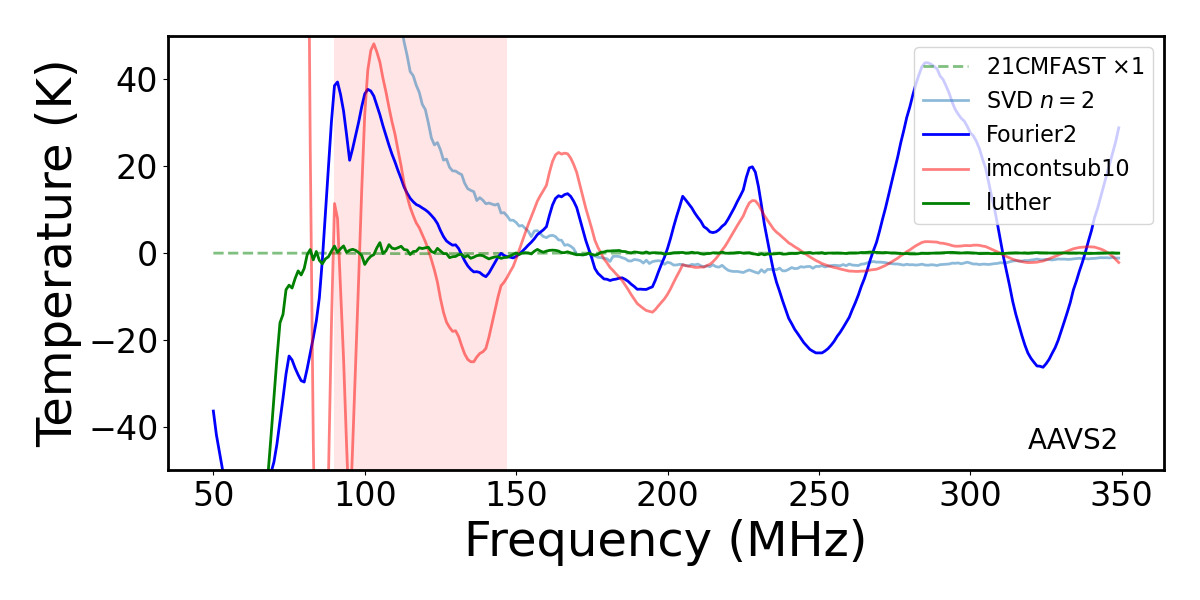}
    \includegraphics[width=0.48\textwidth]{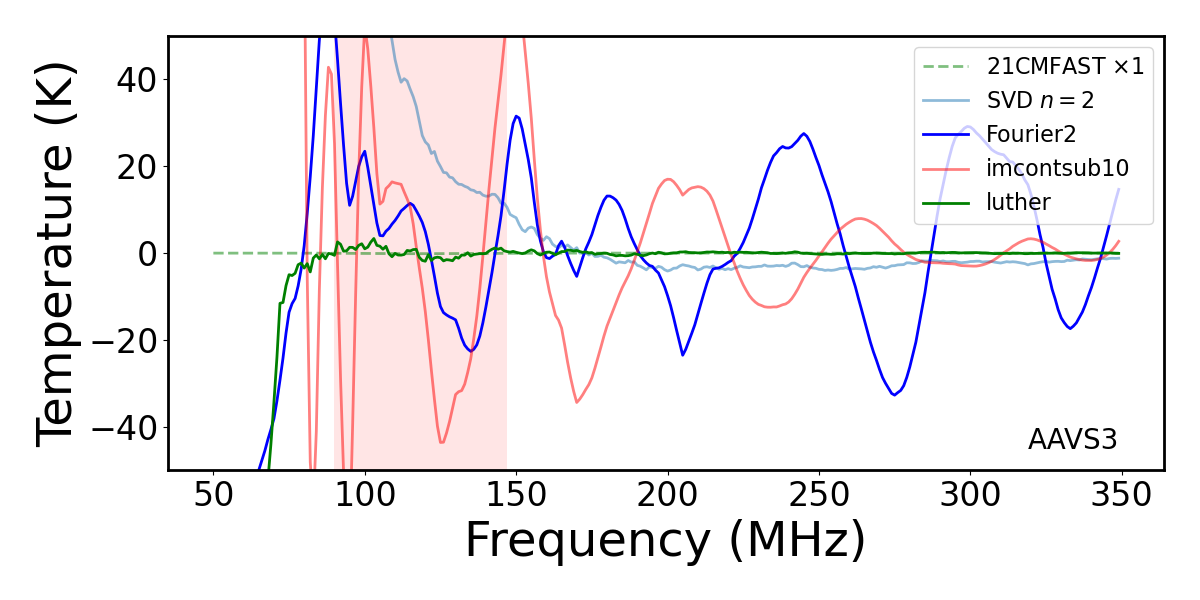}
    \includegraphics[width=0.48\textwidth]{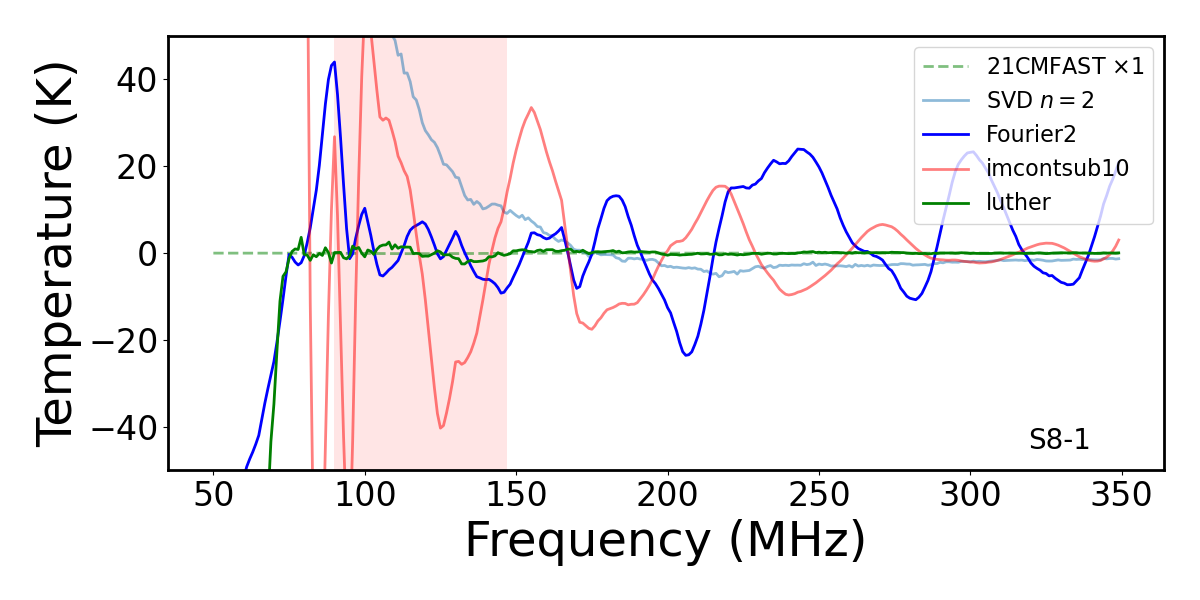}
    \includegraphics[width=0.48\textwidth]{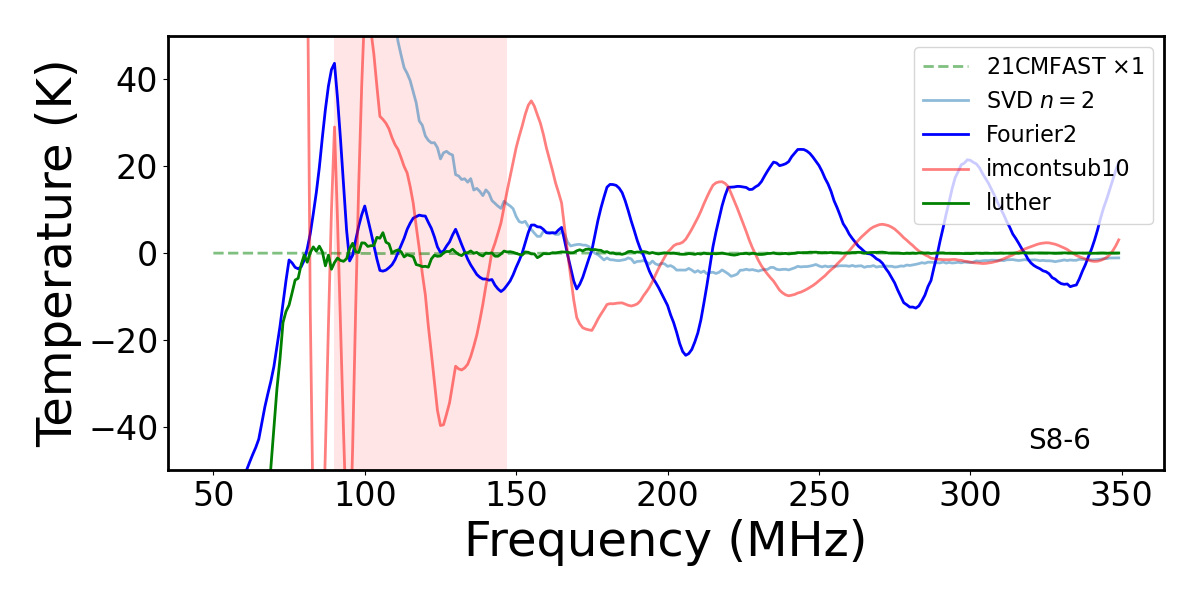}
    \caption{The result of common analysis techniques to arrive at a `reduced' drift-scan spectrum by continuum subtraction, flattening, and inverse reddening of the AAVS2, AAVS3, S8-1 and S8-6  waterfall plots in Figure~\ref{fig:drift}. The spectrum labelled `SVD' has been subject to treatment by Singular Value Decomposition with $n=2$ singular values removed. The spectrum labelled `Fourier' is the time-average of the spectra in the quietest half of the sky followed by removal of the 2 strongest Fourier components. The spectrum labelled `{\tt imcontsub}' is also the time-average of the spectra in the quietest half of the sky, but followed by removal of a polynomial of degree 10. The spectrum labelled `{\tt luther}' has been subject to strong-source bandpass removal and subtraction of a polynomial of degree 3. None of the spectra is able to recover the artificial Cosmic Dawn signal (labelled `21CMFAST') shown by the green dashed line, whose width is indicated by the light red shaded area.}
    \label{fig:driftspec}
\end{figure*}

\begin{figure*}
    \centering
    \includegraphics[width=0.48\textwidth]{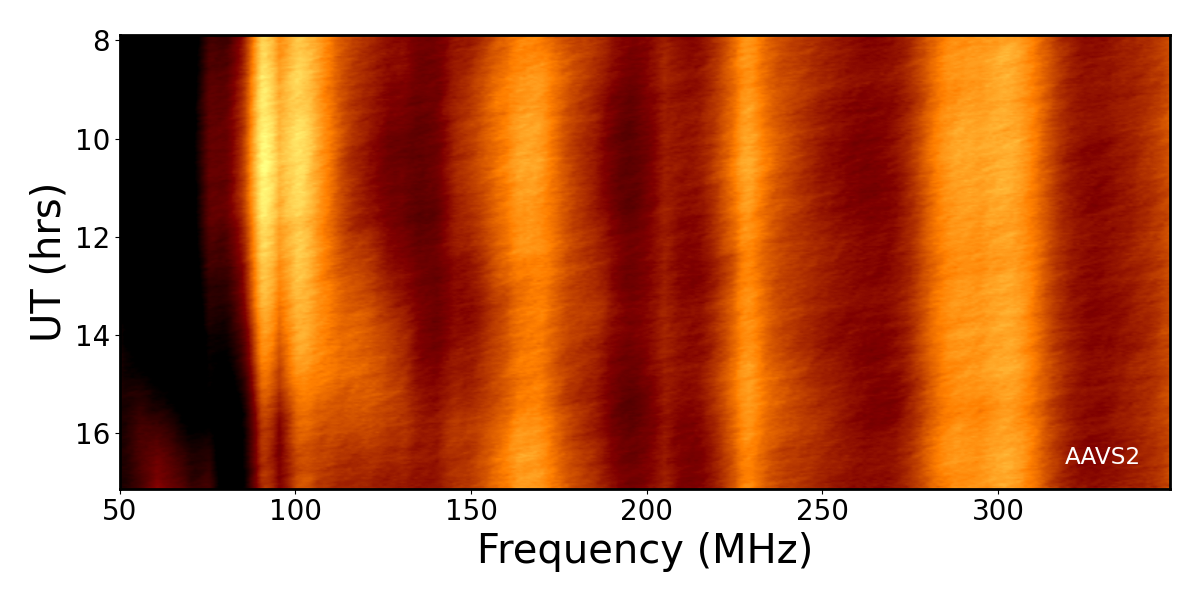}
    \includegraphics[width=0.48\textwidth]{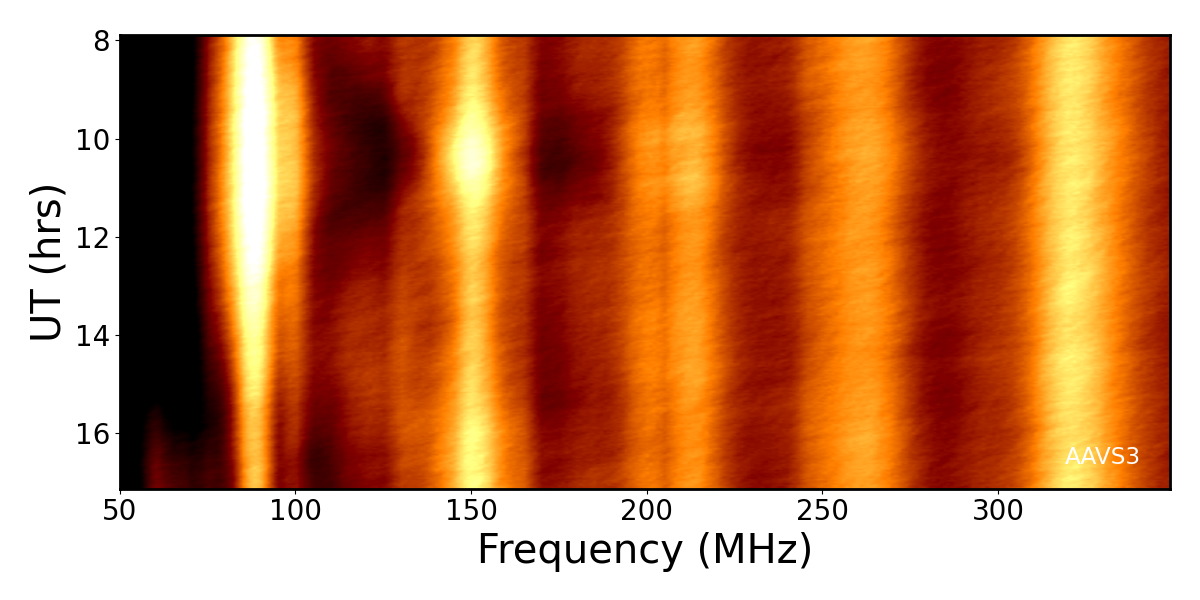}
    \includegraphics[width=0.48\textwidth]{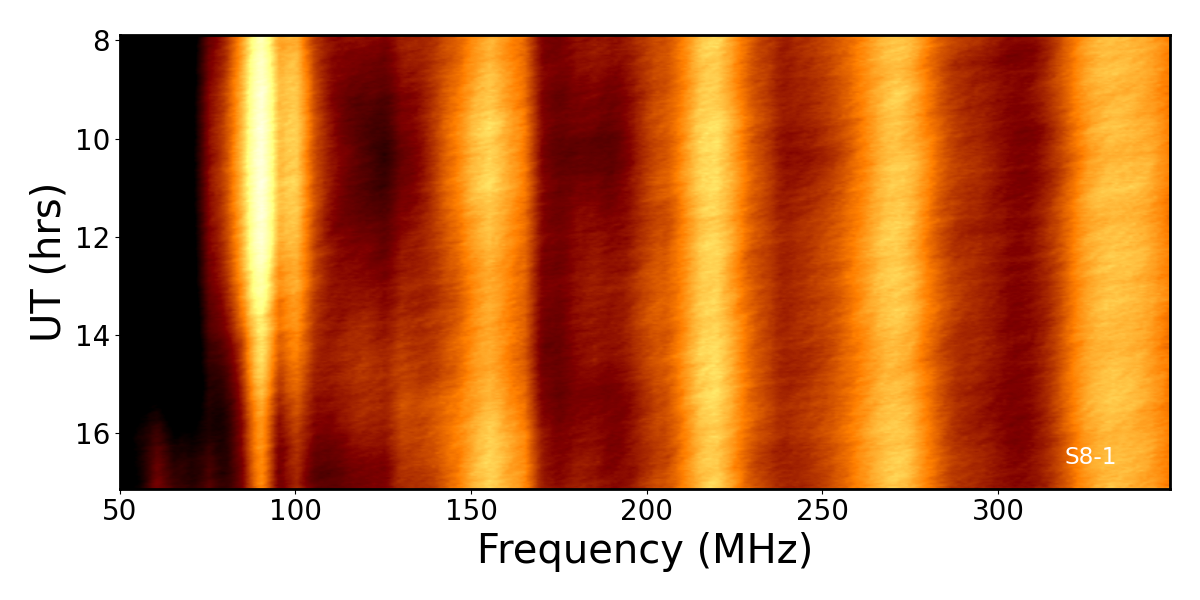}
    \includegraphics[width=0.48\textwidth]{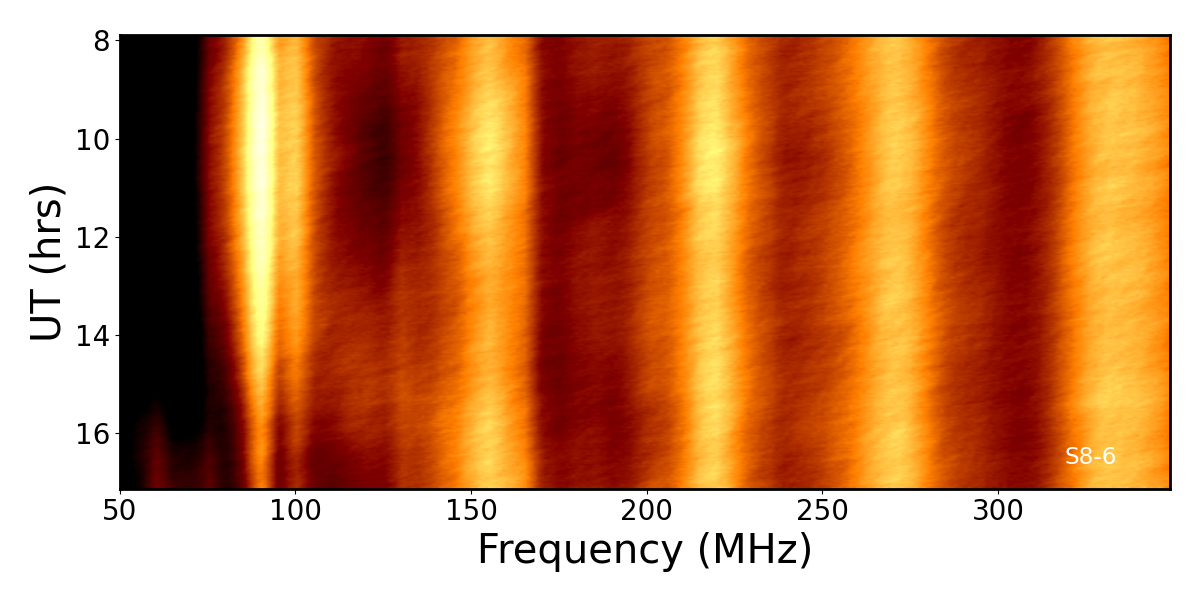}
    \caption{Simulated Stokes $I$ tracks of the South Galactic Pole (frequency in MHz vs time in hrs) for four SKA-Low station configurations. The Low Frequency Sky Model of \citet{2017MNRAS.469.4537D} has been multiplied by the position-dependent (i.e. Azimuth and Zenith Angle) spectral gain model for the following configurations: (top left) AAVS2; (top right) AAVS3; (bottom left) S8-1; and (bottom right) S8-6. For clarity, the spectra have been de-reddened assuming a uniform spectral index of $-2.7$ and a normalisation frequency of 160 MHz. An artificial cosmic dawn signal has been added, but is too weak to be seen here. Noise has been added as described in the text. All the structure in the image is from gain fluctuations in the aperture plane as a result of intra-station antenna interactions. The simulation is based on an SGP track on 1 January 2025. The intensity range is 170 K to 320 K (linear scale).}
    \label{fig:track}
\end{figure*}

\begin{figure*}
    \centering
    \includegraphics[width=0.48\textwidth]{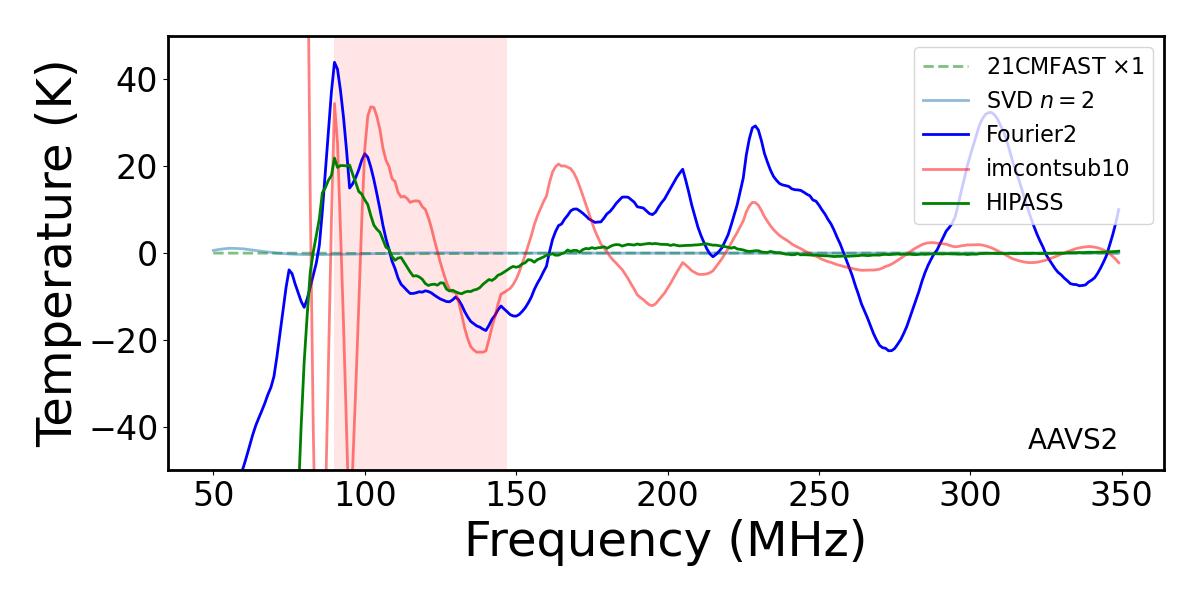}
    \includegraphics[width=0.48\textwidth]{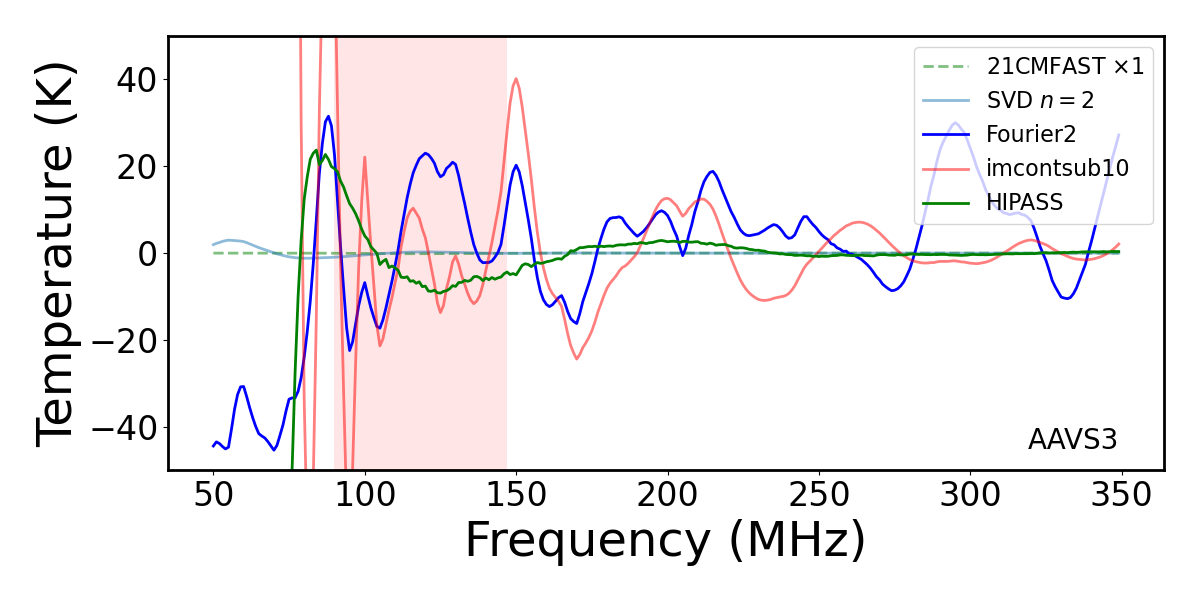}
    \includegraphics[width=0.48\textwidth]{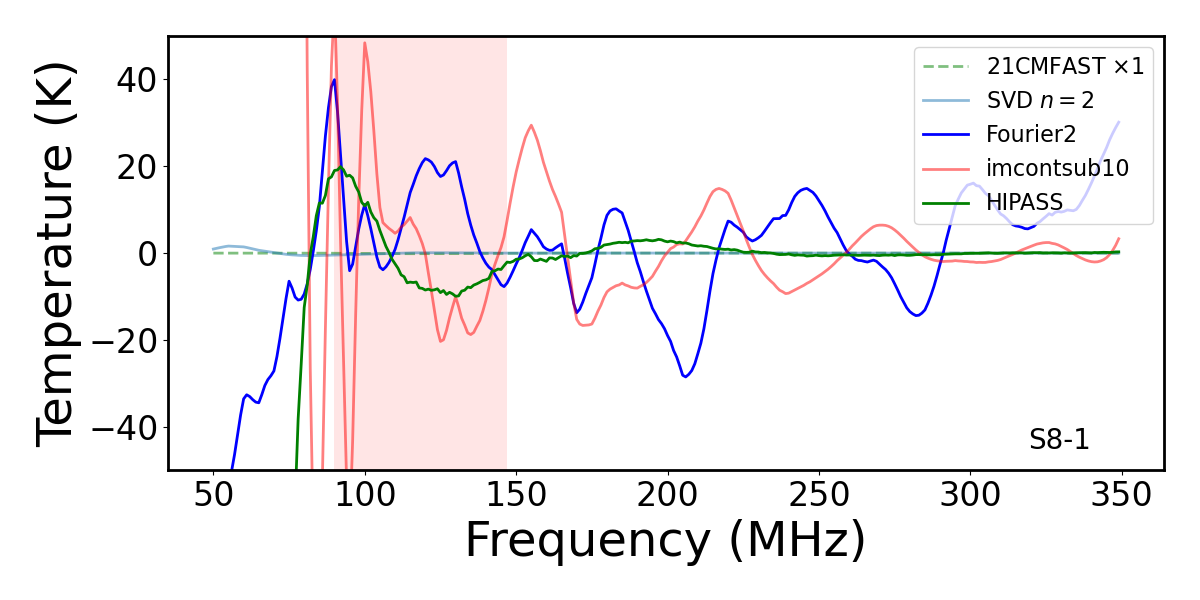}
    \includegraphics[width=0.48\textwidth]{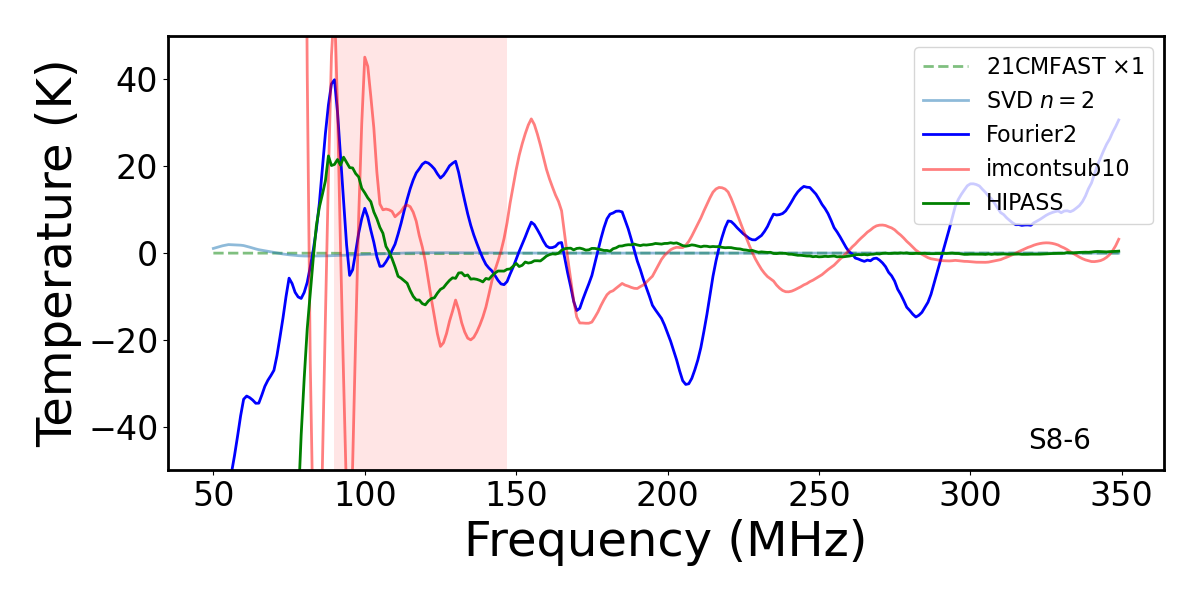}
    \caption{The result of common analysis techniques to arrive at a `reduced' track spectrum by continuum subtraction, flattening, and inverse reddening of the AAVS2, AAVS3, S8-1 and S8-6  waterfall plots in Figure~\ref{fig:track}. The spectrum labelled `SVD' has been subject to treatment by Singular Value Decomposition with $n=2$ singular values removed. The spectrum labelled `Fourier' is the time-average of all spectra, followed by removal of the 2 strongest Fourier components. The spectrum labelled `{\tt imcontsub}' is also the time-average of all spectra, but followed by removal of a polynomial of degree 10. The spectrum labelled `{\tt HIPASS}' is the mean of the middle 50\% (in time) of spectra, bandpass calibrated by the mean of the remaining spectra and subtraction of a polynomial of degree 3. None of the spectra is able to recover the artificial Cosmic Dawn signal (labelled `21CMFAST') shown by the green dashed line, whose width is indicated by the light red shaded area.}
    \label{fig:trackspec}
\end{figure*}

\subsection{Drift scan}
Drift scans, where the sky is allowed to rotate across the telescope field of view, is a well-known method in optical and radio astronomy for minimising variations in instrumental calibration. It widely used in low-frequency continuum and spectral-line observations \citep{2015PASA...32...25W, 2023ApJ...954..139L} as it limits chromatic effects in gain and beam shape as discussed here in the context of SKA stations (see Section~\ref{sec:spectral}).

In order to simulate a drift scan, we calculated the spectral response and station beam of SKA-Low (and prototype) stations at the zenith, and multiplied this into the LFSS sky model and Cosmic Dawn signal in intervals of 3.6 min over a 24-hr period. This covers a track of sky at a Decl.\ of approximately $-27$ deg. The thermal and $1/f$ noise products were also folded in, and $T_{\rm rx}$ added.

The resultant frequency vs. time waterfalls are shown for four SKA-Low station configurations in Figure~\ref{fig:drift}: AAVS2, AAVS3, S8-1 and S8-6. 
S8-1 and S8-6 are identical but with different rotations. Due to the strong variation across the frequency range (a factor of $\sim 130$), Figure~\ref{fig:drift} has been `de-reddened' using a constant power-law index of 2.7 in temperature (corresponding to 0.7 in flux density).

The structure in Figure~\ref{fig:drift} is dominated by horizontal stripes resulting from the passage of sources (e.g. the Galactic Plane) through the zenith. However, strong vertical stripes are also seen -- these are the result of the spectral gain modulation from antenna interactions and reflect the gain modulation in Figure~\ref{fig:power-spectrum-ground}.

Four different analysis techniques have then been used to `reduce' the mock dataset as follows :
\begin{enumerate}
    \item Continuum subtraction ({\tt imcontsub}): a mean spectrum was formed using the 50\% of spectra with the lowest median absolute deviation (i.e drawn from the quietest parts of the sky). A polynomial fit (degree 10) was then subtracted.
    \item Fourier decomposition: as above, a mean spectrum was formed from the quiet part of the sky. This was followed by removal of the 2 strongest Fourier components.
    \item Strong-source bandpass correction ({\tt luther}): a reference bandpass was formed by summing and normalising the spectra from all times where the spectrally averaged flux density is above the median. The averaged spectrum from the remaining data has then been corrected by this bandpass, then fit with a polynomial (degree 3).
    \item Singular Value Decomposition (SVD): this technique is commonly used in EOR and intensity mapping experiments to separate foreground signal, instrumental distortion and cosmic signals \citep{2023MNRAS.522.1022S, 2024ApJ...967...87W}. We have removed $n=2$ singular values.
\end{enumerate}

In all cases, the final spectrum was then `re-reddened' to recover the correct flux density scale. 

As seen in Figure~\ref{fig:driftspec}, none of the analysis techniques is able to recover the Cosmic Dawn signal from any of the four SKA station configurations. This is unsurprising given the difficulty of separating small all-sky signals from bandpass distortions. Although there are detailed differences in the chromatic response of the derived spectra using the four station configurations, there are no winners due to the fact that the spectral response (apart from $1/f$ noise) is invariant with time when drift scanning.

In terms of the spectral variance of various analysis techniques, there is a much greater difference. The bespoke {\tt luther} method \citep{2016AJ....151...52S} seems to perform best. In terms of signal loss, assessed using artificially high Cosmic Dawn signals, {\tt luther} also does well (only $\sim 50$\% signal loss), but polynomials with $n>5$ oversubtract the Cosmic Dawn signal, as do SVD methods with $n>1$.
However, none of the analysis techniques were able to approach the threshold imposed by $1/f$ and thermal noise due to the combination of station chromaticity and spectral variance in the LFSM sky model. 
The `typical' rms deviations of the `mock-reduced' spectra are listed in Table~\ref{tab:driftrms} ($1/f$ noise causes slight variance in the rms values). The highest $|S/N|$ ratio is 0.1 (AAVS2/{\tt luther}), and the lowest is $1.6\times10^{-4}$ (AASV3/{\tt SVD} $n=1$). The AAVS3 configuration generally has higher rms.

\begin{table}
    \centering
    \begin{tabular}{lcccc}
    \hline
    Method   & \multicolumn{4}{c}{Array}   \\
             & AAVS2 & AAVS3 & S8-1 & S8-6  \\
    \hline
    {\tt imcontsub} ($n=10$) & 25.2K & 29.7K & 29.3K &  29.8K \\ 
    Fourier ($n=2$)          & 14.5K & 16.4K &  9.6K &  9.9K \\ 
    {\tt luther}             &  0.9K &  1.2K &  1.2K &  1.3K \\ 
    SVD ($n=1$)              & 519.0K & 549.0K & 537.4K & 534.1K \\ 
    SVD ($n=2$)              & 41.7K & 44.5K & 43.8K &  43.2K \\ 
    \hline
   
    \end{tabular}
    \caption{Typical rms deviations of the `mock-reduced' drift-scan spectra from the Cosmic Dawn spectrum within the 20\% window shown in Figure~\ref{fig:21cmfast} using various algorithms: polynomial removal ({\tt imcontsub}), Fourier filtering, {\tt luther}, and singular value decomposition (SVD). The mean values for the sky and Cosmic Dawn temperatures are 1002~K  and $-0.088$~K in the same window.}
    \label{tab:driftrms}
\end{table}

\begin{table}
    \centering
    \begin{tabular}{lcccc}
    \hline
    Method   & \multicolumn{4}{c}{Array}   \\
             & AAVS2 & AAVS3 & S8-1 & S8-6  \\
    \hline
    {\tt imcontsub} ($n=10$) & 21.6K & 20.0K & 21.9K &  22.9K \\ 
    Fourier ($n=2$)          & 16.5K & 13.9K & 10.8K &  10.7K \\ 
    {\tt hipass}             & 9.9K & 7.6K & 9.8K &  10.7K \\ 
    SVD ($n=1$)              & 273.8K & 276.4K & 278.5K &  278.1K \\ 
    SVD ($n=2$)              & 0.1K & 0.4K & 0.2K &  0.2K \\ 
    \hline
   
    \end{tabular}
    \caption{Typical rms deviations of the `mock-reduced' track spectra from the Cosmic Dawn signal within the 20\% window shown in Figure~\ref{fig:21cmfast} using the same algorithms as for Table~\ref{tab:driftrms}, with {\tt luther} replaced by {\tt hipass}. The low rms of the SVD ($n=2$) method is accompanied by 100\% signal loss. The mean values for the sky and Cosmic Dawn temperatures are 570~K  and $-0.088$~K in the same window.}
    \label{tab:trackrms}
\end{table}

\subsection{Track}

Longer integration times at specific pointings can be achieved by shifting the phase centre of the SKA-Low station to follow a source as the sky rotates. This is the more common observing method. In the case of SKA-Low, this  exposes the user to a position and frequency-variable gain factor discussed in Section~\ref{sec:spectral}. However, it allows quieter parts of the sky to be observed for longer.

We have therefore calculated the Stokes $I$ station gain and beam at all frequencies and sky positions (in azimuth and zenith angle) for each SKA station. We chose an observing position at the South Galactic Pole (SGP) and simulated a day of observing using similar parameters to the drift scan (sky model, thermal noise, $1/f$ noise, cosmological signal).

The resultant frequency vs. time waterfalls are shown for the four SKA-Low station configurations in Figure~\ref{fig:track}. As previously, the data have been `de-reddened' using a constant power-law index of 2.7 in temperature (corresponding to 0.7 in flux density). Compared to Figure~\ref{fig:drift}, no horizontal striping is seen -- the stations are tracking the SGP. However, the spectral structure of each station and its variation with pointing direction is more visible. SGP transit occurs at approximately 10:30 UT on 1 January, and is close to the zenith at the SKA-Low site. The main common features that can be seen in Figure~\ref{fig:track} are, as discussed in Section~\ref{sec:spectral}, array chromaticity and ground reflections which lead to the $\sim60$ MHz pseudo-periodicity. In addition, gain dropouts at 60--80 MHz (over and above the poor antenna efficiency at these frequencies) and horizon fluctuations at 60--100 MHz can be seen.

There are detailed differences in the amplitude structure in the Stokes $I$ tracks. This is most prominent for the AAVS3 configuration, where the regular Vogel pattern gives rise to a zenith gain peak at 150 MHz due to first-order reflections between antennas. However, second- and higher-order reflections are independent of pointing direction, so these give rise to the stronger vertical strip at the same frequency. 

Similar `analysis' techniques as used for the drift scan have been applied to the track scan, except that strong source bandpass correction (`{\tt luther}') has been replaced by weak source bandpass correction (`{\tt HIPASS}'). In this case, a spectrum was formed by averaging the central 50\% (in time) of all spectra, and bandpass calibrated using the mean of the remaining spectra. As for {\tt luther}, a polynomial of degree 3 was subtracted and the spectrum was re-reddened. The resultant spectra for all analysis methods are shown in Figure~\ref{fig:trackspec}, and the rms deviations from the Cosmic Dawn spectrum are given in Table~\ref{tab:trackrms}. The highest $|S/N|$ ratio is 1 (AAVS2/{\tt SVD} $n=2$), and the lowest is $3\times10^{-4}$ (S8-1/{\tt SVD} $n=1$).

Due the dominance of the antenna standing wave over the array interactions, there is again no significant difference in performance of the Fourier and {\tt imcontsub} techniques for the different array configurations. 
The track results are a slight improvement over drift, implying that dealing with flux and spectral fluctuations in the sky are as challenging as station chromaticity.
The weak-source `{\tt HIPASS}' spectrum has lower rms but, unlike `{\tt luther}', only works for compact spectral sources, and completely subtracts all-sky signals.
The $n=2$ SVD spectrum has the lowest rms (unsurprising, as $\sim 2\times 300$ free parameters have been fit to the time-frequency matrix of data), but this again only works for compact sources in both position and frequency -- i.e.\ the Cosmic Dawn signal is completely subtracted. 
The SVD rms is much lower for the track data due to the fact that the smaller variance in the time dimension lessens the singular value requirement by $\Delta n=1$.

Again, none of the techniques was able to recover the Cosmic Dawn signal, though the overall rms values for the track technique are a slight improvement on the drift scan technique. However,  an important source of  spectral variance not considered here is the frequency-dependence of the embedded element pattern (EEP). In our simulated track, we have only considered a frequency-averaged Stokes $I$ EEP. In practice, the EEP will be frequency dependent, particularly at large zenith angle.

\section{Station beam patterns}

\begin{figure*}
    \centering
    \includegraphics[width=1.0\textwidth]{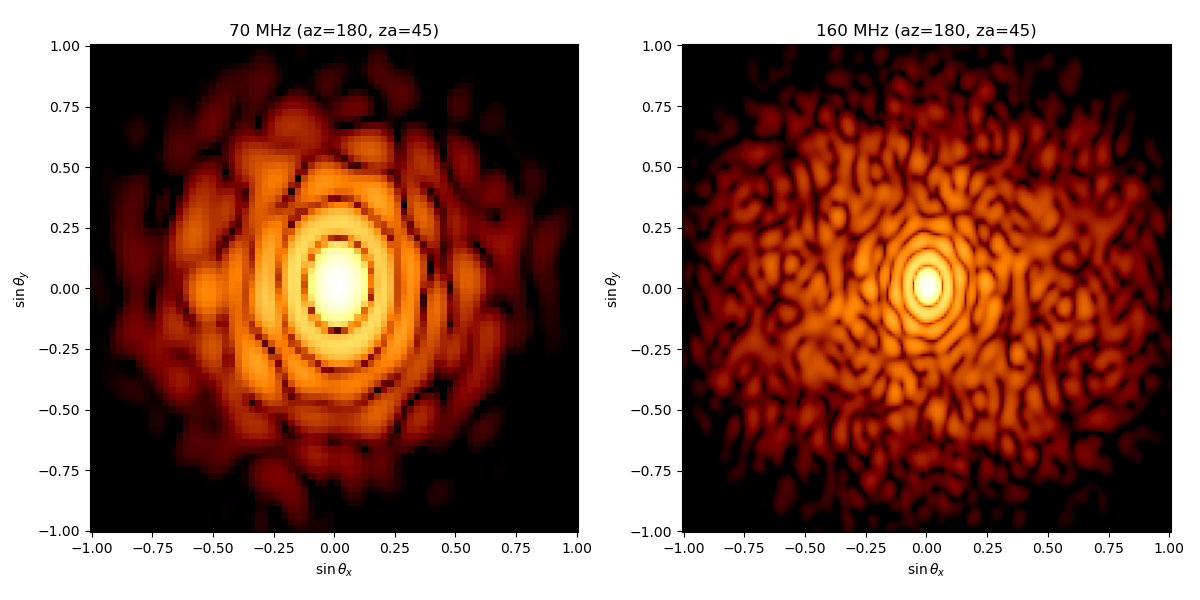}
    \caption{Example SKA-Low S8-1 (power) station beams predicted for (left) 70 MHz and (right) 160 MHz at an azimuth 180 deg, zenith angle 45 deg. The dynamic range is 60dB, and the colour scale is logarithmic. The beams are normalised to unity, and the first positive sidelobe amplitude is $\sim 0.015$ ($-18$dB) of the central peak. No apodisation or primary beam (EEP) correction is applied. }
    \label{fig:beam-pattern}
\end{figure*}

\begin{figure}
    \centering
    \includegraphics[width=1.0\textwidth]{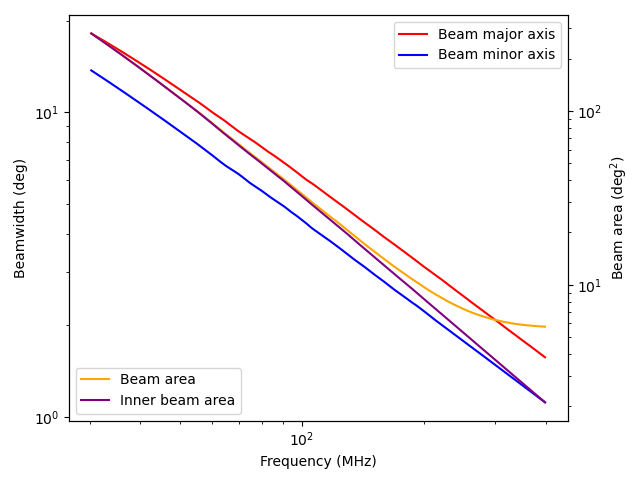}
    \caption{FWHP beamwidth, inner beam area and hemispheric station beam area computed for the SKA-Low S8-1 station.}
    \label{fig:beam}
\end{figure}

Although not the primary aim of this paper, the variation of the station beam with frequency is normally the most problematic behaviour for low-frequency telescopes. It results in mode-mixing of spatial structure in foreground emission with spectral structure \citep{2015PhRvD..91h3514S}. The variation of the beam shape with position on the sky (especially zenith angle) also results in mode-mixing and increases the complexity of Earth rotation synthesis. However, due to the intrinsic flexibility of the SKA-Low aperture arrays, these effects can be somewhat mitigated by apodisation, although this would be at the expense of increasing the spectral variance discussed in Section~\ref{sec:spectral}.

Nevertheless, our analysis allows us to compute the frequency-dependent electric field across the aperture plane as a result of reflections and other mutual coupling. The resultant station beam cube has been formed  by squaring the resultant Fourier transform. Example station beams are shown in Figure~\ref{fig:beam-pattern} at 70 and 160 MHz. There is some spatial variance of the electric field across the SKA-Low station aperture. For example, edge antennas have lower coupling with other antennas. However, this variance seems fairly low, with typical phase deviations across the aperture being only a few degrees. The station beam and sidelobes are therefore fairly well behaved, except at higher frequencies where the station elements become sparsely distributed. 

The frequency-dependent behaviour of the station beamwidth and beam area is shown in Figure~\ref{fig:beam}. The beam major and minor axes are from Gaussian fits to the central lobe of the station beam. The spectral gain features discussed in Section~\ref{sec:spectral} largely affect the beam normalisation and less so the beam pattern. However, as discussed above, there are higher levels of far sidelobes at higher frequencies which result in a substantial increase in the total beam solid angle on the sky.

In addition, there will be other beam pattern effects not studied here which may affect spectral-line observations:

\begin{itemize}
    \item Due to baseline foreshortening, there is considerable spectral variance towards the horizon for SKA stations. Although masked by the ground-plane ripple, some of this can be seen by at late UTs and low frequencies in Figure~\ref{fig:track}. This will not be strongly suppressed by the antenna pattern for azimuths which are at $\pm 90$ deg from the $X$ or $Y$ polarisation axis. The resultant beam pattern will therefore be highly polarised at large zenith angles. As we have only used a Stokes $I$ sky model, the exact consequence of strong polarised sources such as the Galactic Plane, strong radio galaxies and the Sun will need to be the subject of further study, but will result in strong levels of mode-mixing as discussed above. For interferometry, this is mitigated by station rotation \citep{2009wska.confE..17C}

    \item We have applied only a simple ground-plane model in this analysis -- one in which the ground delay is independent of pointing direction (unlike the delays due to antenna configuration). As with the brute force EM models \citep{2022JATIS...8a1017B}, we have also assumed no diffraction effects from the edge of the ground screen.
\end{itemize}

These effects (beam patterns, mode mixing, polarisation, sidelobes, ground plane ripple) require further study and analysis, especially after SKA-Low commissioning data becomes available for comparison with the ideal electromagnetic models and IRF parameterisation methods discussed here.

\section{Further discussion}

We have demonstrated in this paper the utility of the IRF in efficiently computing the effect of element configurations within SKA-Low stations once estimates of some basic antenna properties, particularly reflection coefficients and scattering matrices are known, even in the common low-frequency case case, where they are frequency dependent. This efficiency arises because the location of features in frequency space is largely determined by the various geometric delays between array elements, and therefore their frequency-dependent phase terms. Only the relative amplitude of the features is being determined by the less-well  measured/calculated element scattering properties.

We have shown that the relatively high reflection coefficient of the SKALA4.1 antennas results in inevitable and substantial chromatic behaviour of all SKA-Low station designs, particularly in the EOR frequency range. This chromaticity affects the complex gain (amplitude and phase) of SKA-Low stations. Additive chromaticity (where amplifier noise is scattered between array elements) is likely to be small over the whole SKA-Low band due to the dominance of the sky background over receiver noise.

The analysis presented here suggests that the chromatic behaviour of SKA-Low stations, the variable spectral behaviour of the sky in combination with a simple model of $1/f$ noise will make it difficult to detect faint spectral emission and absorption signatures in station or `autocorrelation' mode. Although these considerations don't apply in normal SKA cross-correlation mode, it should be noted that station complex gains are multiplicative, so a cross-correlation between stations S8-1 and S8-6 will involve a multiplication of their complex spectral gains and a spatial multiplication of their voltage beams. It is emphasised that, although the S8-1 and S8-6 configurations are identical, they are rotated with respect to each other, so their spectral and spatial response have different sky dependence. Even across the station beam, their phase response will not cancel as is normally the case with interferometers with identical elements.
The response of the SKA-Low station clusters, and the SKA-Low core, to faint HI fluctuations (EOR/Intensity mapping) is certainly something that would benefit from further exploration using the IRF technique as such analysis is currently beyond the reach of brute-force electromagnetic simulations. 

An important area of general concern for low-frequency arrays is the presence of strong Galactic and extragalactic foreground emission. The spectral index of the Galactic emission is approximately $-0.7$ in flux density. However, it is usually quoted as $-2.7$ in temperature units because the normal radio telescope beam area scales with the same frequency dependence as the Rayleigh-Jeans relation between brightness temperature $T_B$, and flux density $S_{\nu}$. This spectrally and spatially variant sky results in substantial mode-mixing, where variations in the sky away from the region of interest masquerade as fluctuations in frequency space. Examples of problem areas and mitigation strategies include the following:

\begin{itemize}
    \item Station beam: we have confirmed that the SKA-Low station beam response is likely to be relatively well behaved -- our model predicts little phase and amplitude variation across the aperture plane and predicts that the far sidelobe level will be small (see Figure~\ref{fig:beam-pattern}). However, the main beam area and the sidelobe response is hugely frequency dependent (see Figure~\ref{fig:beam}), especially at frequencies where the array becomes sparse. An obvious mitigation strategy here is station apodisation, which is something that SKA-Low can achieve that no other low-frequency array can achieve. Apodisation, where outer station antennas are down-weighted can cut sidelobe levels and, over a limited frequency range, allow frequency-independence of the station beam shape. For 256 elements in each array, a probable maximum frequency ratio is 2:1. In other words 256 elements at the lowest frequency and 64 effective elements at the highest. This range can be further extended by forming multiple overlapping sub-arrays at high frequencies. The limitations in terms of spectral variance when small numbers of antennas are used to form sub-station beams have not been modelled here, although some of the potential limitations can be seen in Figure~\ref{fig:power-spectrum} where large deviations can be seen in individual element frequency response due to reflections/coupling.

    \item Element pattern: the actual station response is a produce of the station beam considered above and the element beam (or average EEP). Since the SKALA4.1 antenna elements are modified dipoles, there is substantial gain at sky azimuths perpendicular to the polarisation plane, and substantial loss of gain at 90 deg to this . This creates opportunities for mode-mixing with large swathes of the sky such as the Galactic plane, where polarisation levels may be significant. Although we haven't explicitly calculated polarisation power spectra in this analysis, the cross-polarisation $XX$ and $YY$ scattering matrices are well understood, and the delay geometry is identical, with the cross-phase also being dependent on $\sin(2\phi)$, where $\phi$ is the baseline azimuth (or offset azimuth, for rotated stations). As discussed above, apodisation helps here by reducing far-sidelobe sensitivity. In interferometry mode, different pseudo-random configurations would also help -- sidelobes in different locations would not cross-correlate. Station rotation, as implemented within SKA-Low station clusters may also assist.
\end{itemize}

We have not given consideration to interactions between cluster stations. That is also an area for possible future study The longest baseline in SKA-Low station S8 is the S8-1--S8-6 pair considered here. The baseline length is 122 m, or a maximum element separation of 160 m. At low frequencies, where mutual interactions may still be relevant, this could result in frequency structure at in the station gain at $\Delta \nu \approx 2$ MHz. Reverberation, or multiple reflections, will also be a factor below 100 MHz.

\section{Conclusions}
We have conducted a study of the likely levels of chromaticity in SKA-Low stations at frequency resolutions higher than normally accessible to brute-force electromagnetic analyses. 
Our study has reached the following conclusions:

\begin{enumerate}
    \item Variation in spectral gain, due to the station properties alone, will be at the level of $\sim 20-40$\%  in amplitude and $\sim 10-20$ deg in phase, and a factor of $\sim 30$\% higher for individual elements. The effects are worse at the critical frequency, where the observing wavelength is close to the average antenna spacing, and at frequency below 90 MHz, where the element reflection coefficient is high. These effects are in addition to the purely near-field mutual coupling effects  which closely-spaced elements have on each other, and to the fine structure imposed by the frequency-dependent power transmission coefficient of the elements \citep{2020IOJAP...1..253B}. The effects are worse for regular arrays such as the prototype AAVS3 Vogel array or HERA/CHIME etc.\ than for pseudo-random arrays of SKA-Low.
    
    \item Within a finite frequency bands, the impulse response function (IRF), has been shown to be a computational efficient method to calculate complex station gains. Compared to computationally-intensive methods, it predicts the frequency-dependent phase of scattered co-polar and cross-polar signals with reasonable accuracy over the whole SKA-Low band, and is able to predict the amplitude and phase dependence over the station aperture, and therefore the position-dependent extended sidelobe pattern across the sky (a 5-dimensional hypercube, or 6 dimensions if the polarimetric response is also calculated) in a manner which is more efficient than brute-force $m$-mode analysis.

    \item Due to the combined effects of chromaticity (in spectral response and beam shape), bright sky background and $1/f$ noise, SKA-Low stations (individually) will likely be unable to detect signals of similar strength and width to the Cosmic Dawn all-sky signal. It appears only possible to achieve very low $S/N$ ratios (in the range 1 to $10^{-3.8}$) with the observing and reduction techniques investigated here. In fact, signal loss for any all-sky or non-compact signal will make even these low $S/N$ ratios impossible to achieve. This is because it is not straightforward to separate bandpass effects from a spatially-invariant sky signal in either of the observing methods considered here (i.e.\ drift and tracking scans). However, we emphasise that it is not a science goal of the SKA to detect the Cosmic Dawn all-sky signal in station autocorrelation mode. Nevertheless, the gain variations modelled here, especially their dependence on time and position, still apply to cross-correlation gains and will also need to be taken into account when designing cosmological spectral-line interferometric experiments.

    \item If it is a technical possibility, short-baseline SKA-Low experiments designed to detect faint spectral features should consider apodising the aperture function at the beamformer level so that: (a) distant sidelobes are suppressed, and (b) that a frequency-dependent beam can be created over the spectral range of interest.
\end{enumerate}
\begin{acknowledgement}
We thank David Davidson, Mike Kriele and Ravi Subrahmanyan and the SKAO for discussions and providing access to electromagnetic modelling data.
\end{acknowledgement}

\bibliography{AAVS.bib}

\end{document}